\documentclass[journal,draftclsnofoot,onecolumn,12pt,twoside]{IEEEtranTCOM}
\normalsize

\usepackage{pgfplots}

\ifCLASSINFOpdf
\else
\fi

\usepackage{amssymb}
\usepackage{amsthm}
\usepackage[cmex10]{amsmath}
\usepackage{algorithm}
\usepackage{algorithmic}
\usepackage{array}
\usepackage{mdwmath}
\usepackage{mdwtab}
\usepackage[tight,footnotesize]{subfigure}
\usepackage{url}
\usepackage[center]{caption}
\usepackage{float}
\usepackage{subfigure}
\usepackage{epsfig,mathptm, verbatim}
\usepackage{graphicx}
\usepackage{amssymb}
\usepackage{tikz}
\usepackage{hyperref}
\usepackage{breakurl}
\usepackage{amsbsy}
\renewcommand{\vec}[1]{\mathbf{#1}}
\newcommand{\sfrac}[2]{\raise0.1ex\hbox{$#1$} \! \! \left/ \! \lower0.6ex\hbox{$#2$}\right.}
\newcommand{\sfracsmall}[2]{\raise0.1ex\hbox{$#1$} \! \! \left/ \! \lower0.6ex\hbox{$#2$}\right.}

\theoremstyle{plain}
    \newtheorem{theorem}{Theorem}
    \newtheorem{corollary}{Corollary}[theorem]
    \newtheorem{lemma}{Lemma}
	\newtheorem{remark}{Remark}

    \newtheoremstyle{TheoremNum}
        {\topsep}{\topsep}              
        {\itshape}                      
        {}                              
        {\bfseries}                     
        {.}                             
        { }                             
        {\thmname{#1}\thmnote{ \bfseries #3}}
    \theoremstyle{TheoremNum}

\usepackage{cite}
\usepackage[square, comma, sort&compress,numbers]{natbib}
\newcolumntype{L}[1]{>{\raggedright\let\newline\\\arraybackslash\hspace{0pt}}m{#1}}
\newcolumntype{C}[1]{>{\centering\let\newline\\\arraybackslash\hspace{0pt}}m{#1}}
\newcolumntype{R}[1]{>{\raggedleft\let\newline\\\arraybackslash\hspace{0pt}}m{#1}}
\usepackage{multirow}

\date{}

\usepackage{lipsum}

  \hypersetup{linktocpage} 
  \hypersetup{
    colorlinks,
    citecolor=blue,
    filecolor=black,
    linkcolor=red,
    urlcolor=black
}

\begin{document}
\ifCLASSOPTIONtwocolumn
\title{Fractional Interference Alignment: An Interference Alignment Scheme for Finite Alphabet Signals}
\else
\title{Fractional Interference Alignment: An Interference Alignment Scheme for Finite Alphabet Signals}
\fi
\ifCLASSOPTIONtwocolumn
\author{B Hari Ram \hspace{4mm} K Giridhar \\
Department of Electrical Engineering \\
 Indian Institute of Technology Madras \\
  Chennai-600036, India\\
  Email:[hariram, giri]@tenet.res.in\\
  \vspace{-12mm}
  }
\else
\vspace{-4mm}
\author{B Hari Ram \hspace{4mm} K Giridhar \\
\vspace{-5mm}
Department of Electrical Engineering \\
\vspace{-5mm}
 Indian Institute of Technology Madras \\
\vspace{-5mm}
  Chennai-600036, India\\
\vspace{-5mm}
  Email:[hariram, giri]@tenet.res.in\\
  \vspace{-20mm}
}
\fi
\maketitle

\begin{abstract}
Interference Alignment (IA) is a transmission scheme which achieves $\sfrac{1}{2}$ Degrees-of-Freedom (DoF) per transmit-antenna per user. The constraints imposed on the scheme are based on the linear receiver since conventional IA assumes Gaussian signaling. However, when the transmitters employ Finite Alphabet (FA) signaling, neither the conventional IA precoders nor the linear receiver are optimal structures. Therefore, a novel Fractional Interference Alignment (FIA) scheme is introduced when FA signals are used, where the alignment constraints are now based on the non-linear, minimum distance (MD) detector. Since DoF is defined only as signal-to-noise ratio tends to infinity, we introduce a new metric called SpAC (number of Symbols transmitted-per-transmit Antenna-per-Channel use) for analyzing the FIA scheme. The maximum SpAC is one, and the FIA achieves any value of SpAC in the range $[0,1]$. 
The key motivation for this work is that numerical simulations with FA signals and MD detector for fixed SpAC (=$\sfrac{1}{2}$, as in IA) over a set of optimization problems, like minimizing bit error rate or maximizing the mutual information, achieves a significantly better error rate performance when compared to the existing algorithms that minimize mean square error or maximize signal-to-interference plus noise ratio. 

\end{abstract}


\section{Introduction} \label{sec:Intro}
A wireless network with $K$ number of transmitters and receivers, forming $K$ transmitter-receiver (Tx-Rx) pairs, is called a $K-$user Interference Channel (IC). In a $K-$user IC, each transmitter has a useful message to only its paired receiver, and the presence of other co-channel transmitters in the vicinity results in $K-1$ interfering terms getting added to the desired signal at each of the $K$ receivers.
The IC was first introduced by Shannon in \cite{Shannon1961}, and studied further by Ahlswede \cite{Ahlswede1974}. The capacity of the IC has been studied \cite{Carleial1978,Sato1981,Han1981, Gamal1982,Costa1985}, and in \cite{Etkin2008}, a tighter outer bound within 1 bit/Hz/second from the channel capacity has been described. Finding the capacity region even for the $2-$user IC remains an open problem.

In \cite{Cadambe2008}, Degrees of Freedom (DoF) for the $K-$user IC was studied. DoF represents the scalar pre-multiple of the logarithmic term in the capacity expression as signal-to-noise ratio (SNR) tends to infinity. An Interference Alignment (IA) scheme was introduced in \cite{Cadambe2008}, where the precoders are designed such that the DoF is maximized in a $K-$user IC. However, a closed form solution was provided for only the $K=3$ user IC, which achieves a total DoF of $\sfrac{3M}{2}$, where $M$ represents the number of antennas at all nodes (transmitter and receiver). For a general $K-$user IC, the total DoF of $\sfrac{KM}{2}$ was achieved asymptotically as symbol extension factor (SEF)  tends to infinity \cite{Cadambe2008}. 
Hence, in \cite{Gomadam2008, Peters2009,Shen2010, Schmidt2009, Papailiopoulos2012}, various iterative algorithms have been introduced to obtain optimal precoders, with the dimension of the precoder matrices fixed as per the dimension of the IA solution. Numerical results were provided in these papers, showing that as noise variance  tends to zero, the performance of the optimization problems converges to the performance of the IA solution. 

In practical digital communications, the transmitter uses Finite Alphabet (FA) signals. 
However, the conventional IA solution \cite{Cadambe2008} as well as the iterative algorithms in \cite{Gomadam2008, Peters2009,Shen2010, Schmidt2009, Papailiopoulos2012}, assume the usage of Gaussian signaling at all the transmitters. 
The main focus of this work is to extend the IA design, in a more careful manner to the case when all the transmitters use FA signals. In \cite{HariRam2013}, and \cite{Fadlallah2013}, precoders have been designed by maximizing the Mutual Information~(MI) of the FA signal sets. We will refer to such  schemes, which replace the objective functions in \cite{Gomadam2008, Peters2009,Shen2010, Schmidt2009, Papailiopoulos2012}, with functions such as MI or bit error rate~(BER), as Extended IA (EIA) schemes. The EIA is a rather straight forward extension\footnote{Actually, since IA uses $M\times\sfrac{M}{2}$ dimension to the precoder, the EIA also uses only the same dimension, even for FA signals.} of IA when FA signals are used.

In the first part of this work, we consider four different optimization problems with different objective functions which are appropriate for FA signals, namely: (a)~Symbol Error Rate (SER), (b)~BER, (c)~MI for FA signaling, and, (d)~Minimum Distance~(MD) between the constellation points seen at the receiver. For these optimization problems, we will analytically show that aligning the interference within a $\sfrac{M}{2}$ dimensional subspace in the reciprocal channel is nothing but one of the many locally optimal points. 
Since alignment of interference in the reciprocal channel is a locally optimal point, a simple Conjugate Gradient Descent (CGD) is used to obtain sub-optimal precoders, with the precoder dimensions fixed to $M \times \sfrac{M}{2}$. 
The sub-optimal precoders provided by CGD achieve  significant gain in BER performance when compared to the existing algorithms \cite{Gomadam2008, Peters2009,Shen2010, Schmidt2009, Papailiopoulos2012}, when an appropriate non-linear receiver is used (for both EIA scheme and algorithms in \cite{Gomadam2008, Peters2009,Shen2010, Schmidt2009, Papailiopoulos2012}). However, when a linear receiver is employed, the BER performance of EIA scheme floors, indicating that the interference is not aligned within $\sfrac{M}{2}$ dimensions.

Motivated by the fact that the interference is not aligned within $\sfrac{M}{2}$ dimensions (even at infinite SNR), we propose a novel Fractional Interference Alignment (FIA) scheme in the second part of the work. The key differences with respect to the IA scheme \cite{Cadambe2008} are: (i)~In the FIA scheme, each transmitter uses FA signal sets, and (ii)~A non-linear symbol detector is preferred at the receiver. Since non-linear detectors are used at each receiver, the alignment constraint is different for the FIA scheme. The optimal non-linear receiver for FA signals is the MAP or ML (Maximum A-Posteriori, or, Maximum Likelihood) receiver \cite{Willems2010}. Further, when the interference plus noise is Gaussian distributed with a known covariance, the ML receiver reduces to the computationally simpler minimum distance (MD) detector. Throughout this work, the MD detector \cite{Kuchi2011} is used at each receiver. We propose a new alignment constraint with the key property that $\mathcal{S} \nsubseteq \mathcal{I}$ based on the MD detector \cite{Kuchi2011}, where $\mathcal{S}$ represents the desired signal subspace and $\mathcal{I}$ represents the interfering signal subspace at the receiver. 
Also, in the FIA scheme the interfering signals are aligned perfectly, but the desired signal subspace is allowed to overlap with the interfering signal subspace. Since this overlap is allowed, the dimensions of both $\mathcal{S}$ and $\mathcal{I}$ can be increased higher when compared to the IA scheme \cite{Cadambe2008}. Hence, more number of interfering signals can now be handled at each receiver, or alternatively, each transmitter can send more data streams to the intended receiver.

\subsection{Symbols per Antenna per Channel use (SpAC)}
To bring out more clearly the advantage of FIA,
when compared to the EIA schemes \cite{HariRam2013}, \cite{Fadlallah2013}, we introduce here
a new metric abbreviated as SpAC (Symbols per Antenna per Channel use). SpAC represents the number of message streams sent per transmit-Antenna per Channel use. Unlike DoF, the SpAC metric is a more general expression, since the optimum value (for achieving better BER, or MI, or both) of SpAC will be a function of both signal-to-noise ratio (SNR) and signal-to-interference ratio (SIR). It must be mentioned here that the optimum value of (symmetric) SpAC at infinite SNR is the ratio of (symmetric) DoF per user to the total number of dimensions available for transmission. This is based on the fact that DoF represents the maximum number of streams that a transmitter can send as SNR tends to infinity. 
In the EIA precoder design, the precoders are designed with fixed value of SpAC = $\sfrac{1}{2}$. The algorithms in \cite{Gomadam2008, Peters2009,Shen2010, Schmidt2009, Papailiopoulos2012} have $d$ independent columns (per channel use) in the precoder matrices, where $d\leq\frac{M}{2}$\footnote{When each transmitter is equipped with $M$ antennas, and each receiver is equipped with $N$ antennas, the maximum value of $d$ is given as $\sfrac{(M+N)}{K+1}$ \cite{Yetis2009} when no SEF is used. Since the algorithms in \cite{Gomadam2008, Peters2009,Shen2010, Schmidt2009, Papailiopoulos2012} can be used even with SEF (but in a centralized location), with the precoder dimension being $ML\times dL$, the maximum value of $d$ is $\sfrac{M}{2}$ as $L\rightarrow\infty$ \cite{Cadambe2008}.}. Thus, they all will have SpAC$\leq\sfrac{1}{2}$ even when FA signals are used.
When compared to the conventional IA, the FIA allows overlap between desired and interfering signal subspaces. However, it is not a strict constraint; i.e., if the required SpAC $\leq \sfrac{1}{2}$, then the overlapping of the two subspaces is not required and the constraints in FIA can be reduced to that of the constraints in conventional IA scheme. 
From our numerical result, for a SpAC value of $\sfrac{1}{2}$, the EIA precoder designs which  allow overlap give a better BER performance when compared to the precoders in \cite{Gomadam2008, Peters2009,Shen2010, Schmidt2009, Papailiopoulos2012}  which do not allow any overlap at infinite SNR.
When the required SpAC $> \sfrac{1}{2}$, then overlapping (even at infinite SNR) of the two subspaces becomes a necessary condition.

The maximum achievable value of SpAC is one, and it will be shown that FIA scheme achieves 1 SpAC asymptotically as the number of dimensions available for transmission, $M$, tends to infinity. Hence, FIA can be viewed as a collection of precoder designs to achieve different values of SpAC in the full range $[0,1]$. Indeed, we have chosen to call this method as FIA, since we can achieve any fractional value of SpAC between $[0,1]$. 
It must also be clarified that the FIA precoder design with the constraint $\mathcal{S} \nsubseteq \mathcal{I}$ will not achieve DoF$> \sfrac{1}{2}$.
The FIA scheme can be viewed as a careful extension of the conventional IA scheme \cite{Cadambe2008} by allowing the overlap of the interfering signals and desired signal subspaces, i.e., relaxing the constraint $|\mathcal{S}\cup \mathcal{I}| = |\mathcal{S}| + |\mathcal{I}|$ to $\mathcal{S} \nsubseteq \mathcal{I}$. 
It must be clarified here that the EIA scheme in \cite{HariRam2013},\cite{Fadlallah2013} implicitly keep SpAC$=\sfrac{1}{2}$ while handling FA signals using either linear or non-linear receivers for the given objective functions.
Hence, the proposed FIA is a more general and effective framework than EIA, while extending the IA concept to FA signals. 

We describe FIA schemes for both Multi Input Multi Output (MIMO), as well as Single Input Single Output (SISO) ICs. Since SISO IC has only one antenna at the Tx and Rx nodes, precoding of the transmit signal is done across resources which can be either in time and/or frequency. The number of resource elements used for precoding is known as the symbol extension factor (SEF), since the contribution from one symbol is present across many symbol durations. Hence, the number of dimensions available for transmission, $M$, can represent the SEF in SISO IC. For MIMO IC, symbol extension need not be always utilized, and in such a case, the number of antennas at each transmitter represent the number of dimensions available for transmission. When symbol extension is utilized, the number of dimensions available for transmission is given by the product of the number of transmit antennas and the SEF. 

Even though FIA will be shown to achieve values of SpAC higher than the EIA schemes in \cite{HariRam2013}, \cite{Fadlallah2013}, it does not give a clear insight on the performance of FIA. In fact, it is important that this increase in the value of the SpAC should not degrade the BER performance for a fixed rate. Hence, in our earlier work \cite{HariRam2013c}, we have numerically studied the efficacy of the FIA scheme over the EIA scheme. For the proposed FIA, achievable values of SpAC are obtained theoretically in our current work, while the optimum value of SpAC was obtained in \cite{HariRam2013c} using iterative algorithms. 

\subsection{FIA and other schemes for FA signals}
The idea of allowing the interfering signals to overlap with the desired signal has also been employed in the Partial IA and Interference Detection (PIAID) scheme \cite{Huang2011}. The key differences between PIAID and the proposed FIA are: (i)~PIAID scheme aligns the interference from multiple transmitters, but allows those interfering signals which cannot be aligned, to overlap with the desired signal subspace; on the other hand, the FIA aligns all the interfering signals such that the interference subspace is allowed to overlap with the desired signal subspace, (ii)~PIAID \cite{Huang2011} decodes the non-aligned interference symbols, and cancels their influence on the desired signal.
In \cite{Wu2013, Ganesan2012}, it was shown that $1$~SpAC can be achieved when FA signals are used. In \cite{Wu2013},  a simple power allocation was shown to achieve $1$~SpAC, and \cite{Ganesan2012} uses high SNR approximation of the mutual information to show $1$~SpAC is achievable. However, \cite{Wu2013} and \cite{Ganesan2012} needs joint detection of all the transmitters signal at each receiver to achieve $1$~SpAC.

The disadvantage of decoding the interfering signals are: (a)~the channel state information (CSI) corresponding to the interfering signals should be estimated at each receiver, and (b)~the FA signal constellation of all the interfering signals to be jointly decoded should be known to the receiver, and (c) the detection complexity is higher (function of the number of interfering signal that is decoded, and the corresponding constellation size). 
Unlike \cite{Huang2011, Wu2013, Ganesan2012}, the alignment schemes like the conventional IA, the EIA and the proposed FIA scheme, do not decode any of the interfering signals. We therefore, do not consider these schemes as candidates for comparison in this paper. Both EIA and FIA need to only estimate the covariance of the interfering signals at each receiver. Estimating the interference covariance does not need the CSI of the interfering signals, and also typically consumes a significantly smaller overhead when compared to estimating the CSI of the interfering signals. The FA signal constellation need not be signaled or estimated for the proposed FIA scheme.

In summary, our work shows that aligning the interference within $\sfrac{M}{2}$ dimensions is nothing but one of the many locally optimum points for the considered optimization problems. An iterative CGD algorithm is utilized to obtain a sub-optimal precoder for the EIA scheme. Since the iterative algorithm returns only sub-optimal precoders which do not align the interference, but yet provides a better BER performance when compared to existing schemes, the question to be answered is this: ``\textit{Is it necessary to restrict the} SpAC to \textit{$\sfrac{1}{2}$ at each transmitter}?". This question will be answered with the proposed FIA scheme which can achieve SpAC $=1$ asymptotically, as the dimension of transmission tends to infinity. 
The mathematical constraint $\mathcal{S} \nsubseteq \mathcal{I}$ ensures that the BER of the FIA scheme goes to zero as SNR$\rightarrow\infty$, which we refer to as the ``Zero BER" criterion. This zero BER criterion makes the problem non-trivial while achieving any value of SpAC in the range $[0,1]$. Hence, FIA achieves 1 SpAC while satisfying the zero BER criterion, and in the process, yields a better BER performance when compared to the EIA scheme \cite{HariRam2013c}.

\section{System Model} \label{sec:SysMod}
A $K-$user IC model is considered, in which $K$ Tx-Rx pairs are communicating independently. There are no common messages between any two Tx-Rx pairs. Every Tx-Rx pair acts as an interferer to the other Tx-Rx pairs, and in turn experiences co-channel interference from them. Assuming, each transmitter and receiver are equipped with $M$ antennas, the received signal, $\vec{y}_i$, at the $i^{th}$ receiver is given by,
\begin{equation}\label{eqn:sysMod}
\begin{array}{lll}
\vec{y}_i &=& \sum_{j=1}^K \vec{H}_{i,j} \vec{Q}_j \vec{x}_j + \vec{z}_i, \text{ } i=1 \text{ to } K
\end{array},
\end{equation}
where $\vec{z}_i$ represents zero mean white Gaussian noise vector of dimension $ML\times 1$, with covariance $\sigma^2 \vec{I}_{ML}$, $\vec{x}_j$ represents the transmitted vector signal at $j^{th}$ transmitter of dimension $n_j\times 1$, and $\vec{Q}_j$ represents the precoder matrix at $j^{th}$ transmitter of dimension $ML\times n_j$. Here $\vec{H}_{i,j}$ is a block diagonal $ML\times ML$ channel matrix from the $j^{th}$ transmitter to the $i^{th}$ receiver. All non-zero elements of $\vec{H}_{i,j}$ are identically, independently and continuously distributed, and therefore, $\vec{H}_{i,j}$ is non-singular. $L$ represent the symbol extension factor\footnote{Unless mentioned, it is assumed that symbol extension is used in this work.}.

\textit{Notations Used:} 
If $\vec{A}$ represents the channel matrix, then $\vec{A}^H$ is the reciprocal channel matrix. Also, $\text{I}_M$ represents the $M\times M$ identity matrix, rank($\vec{A}$) is the rank of the matrix $\vec{A}$, $\vec{\tau}(\vec{A})$ represents the Frobenius norm of the matrix $\vec{A}$ ($\vec{\tau}(\vec{A})$ = trace($\vec{A}\vec{A}^\text{H}$)), span($\vec{A}$) is the column space of the matrix $\vec{A}$, and $|\text{span}(\vec{A})|$ is the dimension of the column space of $\vec{A}$ or the rank of the matrix $\vec{A}$. With some abuse of notation we represent the span$(\vec{A})$ by $\vec{A}$, and if we use: (a)~$\vec{A}\circeq \vec{B}$, (b)~$\vec{A} (\subset or \subseteq or \nsubseteq) \vec{B} $, then $\vec{A}$ and $\vec{B}$ represent the span of the matrix $\vec{A}$ and $\vec{B}$, respectively. Further, $\mathcal{X}_i$ represents the set containing all possible values of the transmitted symbol vector $\vec{x}_i$ and the elements are assumed to be ordered, $\vec{x}_{i,j}$ represents the $j^{th}$ vector element of the set $\mathcal{X}_i$, {$d_i$} is the collection of all distance metric at $i^{th}$ receiver, and $d_i^{[jk]}$ represents the distance between $\vec{x}_{i,j}$ and $\vec{x}_{i,k}$ ($\vec{x}_{i,j}, \vec{x}_{i,k} \in \mathcal{X}_i$), at the $i^{th}$ receiver. Finally, let $\vec{e}_i$ be the axis vector where the $i^{th}$ element is unity and all other elements are zero.

\section{Optimization Problem} \label{sec:OptProb}
Let $\mathcal{X}_i$ represent the set containing all possible vector symbols. The optimization problem is formulated as the function of the distance measure ($d_i^{[jk]}$), and is given as,
\begin{equation} \label{eqn:Opt_prob}
\begin{array}{ll}
 \operatorname*{Optimize}\limits_{\vec{Q}_j,\text{ j=1 to K}} & C = \sum_{i=1}^K f_i(\{d_i\}|\vec{H}_{i,1},\cdots,\vec{H}_{i,K}) \\
\text{subject to} & \tau(\vec{Q}_i) \leq P_{\text{i}}, \quad i=1\text{ to }K \\
\end{array},
\end{equation}

\noindent where $C$ is the objective function to be optimized. In this work, the optimization in (\ref{eqn:Opt_prob}) will be solved as a minimization problem by introducing a negative sign for maximization problems.
In order to maintain a fairness between the Tx-Rx pairs in the optimization problem, the same objective function is chosen for all the Tx-Rx pairs in all the numerical results, i.e., $f_i(\cdot) = f(\cdot)$. However, it will be shown in section~\ref{sec:extremPts} that even if the objective functions are different, as long as it is a function of the distance measure, the conventional IA scheme will be a locally optimal solution.
The distance measure $d_i^{[jk]}$ is given~by,
\begin{equation}\label{eqn:distMeas}
\begin{array}{lll}
d_i^{[jk]} = (x_i^{[j]}-x_i^{[k]})^H\vec{Q}_i^H \vec{H}_{i,i}^H \vec{R}_i^{-1} \vec{H}_{i,i} \vec{Q}_i (x_i^{[j]}-x_i^{[k]})
\end{array},
\end{equation}
where $\vec{R}_i$ represent the covariance of the interference-plus-noise term, namely
\ifCLASSOPTIONtwocolumn
\begin{equation}\label{eqn:IntCov}
\begin{array}{lll}
\vec{R}_i = \tilde{\vec{R}}_i + \sigma^2 \text{I}_M\\
\tilde{\vec{R}}_i = \sum_{j=1, \text{ } j\neq i}^K \vec{H}_{i,j}\vec{Q}_j \left(\vec{H}_{i,j}\vec{Q}_j\right)^\text{H} 
\end{array},
\end{equation}
\else
\begin{equation}\label{eqn:IntCov}
\begin{array}{lll}
\vec{R}_i = \tilde{\vec{R}}_i + \sigma^2 \text{I}_M; \quad
\tilde{\vec{R}}_i = \sum_{j=1, \text{ } j\neq i}^K \vec{H}_{i,j}\vec{Q}_j \left(\vec{H}_{i,j}\vec{Q}_j\right)^\text{H} 
\end{array},
\end{equation}
\fi
and $\tilde{\vec{R}}_i$ represents the covariance of the interfering signals arriving at the $i^{th}$ receiver. 
\subsubsection*{Examples} 
Some commonly used objective functions are,

\textit{Symbol Error Rate} \cite{Cioffi2007a}: 
\begin{equation} \label{obj:SER}
\begin{array}{ll}
f_{\text{SER}} (\{d_i\}) = \sum_{j,k} \; Q(\frac{1}{\eta}d_i^{[jk]})
\end{array}.
\end{equation}

\textit{Bit Error Rate} \cite{Cioffi2007a}:
\begin{equation} \label{obj:BER}
\begin{array}{l}
f_{\text{BER}} (\{d_i\}) = \sum_{j,k} \; \beta_{jk} \;\; Q(\frac{1}{\eta}d_i^{[jk]})
\end{array}.
\end{equation}

\textit{Mutual Information} \cite{HariRam2013}:
\begin{equation} \label{obj:MI}
\begin{array}{ll}
f_{\text{MI}} (\{d_i\}) = -\sum_{j} \text{log}_2 \sum_{k} exp(-\frac{1}{\eta}d_i^{[jk]})
\end{array}.
\end{equation}

\textit{Minimum Distance} (The minimum distance is replaced by summation \cite{Moon2000}):
\begin{equation} \label{obj:MD}
\begin{array}{ll}
f_{\text{MD}} (\{d_i\}) = \operatorname*{min}\;\;d_i^{[jk]}\;\;\; \operatorname*{=}\limits^{(a)} \;\; (\sum_{j,k} (d_i^{[jk]})^{-r})^{-\frac{1}{r}}, r\rightarrow \infty
\end{array}.
\end{equation}

In (\ref{obj:SER}) thro (\ref{obj:MI}), the constant $\eta = 2$, and $Q(x) = \int_x^\infty \frac{1}{\sqrt{2\pi}} e^{-\frac{u^2}{2}}\; du$. 
Also, $\beta_{jk}$ in \eqref{obj:BER} represents the number of errors when the codeword $\vec{x}_i^{[j]}$ is erroneously detected as $\vec{x}_i^{[k]}$. 
For the minimum distance optimization in \eqref{obj:MD}, $^{(a)}$ is obtained using the approximation from \cite{Moon2000}, $r$ is independent of the precoder and the channel matrices, and $\sum_{j,k} (d_i^{[jk]})^{-r}$ is used for $f_{MD}(\{d_i\})$ in the remainder of the paper.
In all the objective functions in (\ref{obj:SER}) - (\ref{obj:MD}), from the definition $d_i^{[jk]}$, the interference plus noise term is modeled as colored Gaussian noise. Although this is unlikely to be strictly true, it is a good approximation when the number of interfering signals is large \cite{Chiani1997},
(i)~with the 
increase in number of Tx and Rx antennas resulting in an increase in the number of streams sent by each transmitter, or (ii)~simply by increasing $K$, which is the number of Tx-Rx pairs. In general, this represents a good approximation\footnote{The BER performance with colored Gaussian model for interference plus noise approaches the performance of an IC where the interference is actually colored Gaussian, as the number of interfering signal increases. \cite{Chiani1997}} \cite{Chiani1997} except for small values of $M$ and $K$. However, throughout the work, this approximation is followed, regardless of the value for $M$ and $K$. The tightness of the results improves when $M$ or $K$ or both are large.

\section{Interference Alignment: An Optimal Point} \label{sec:extremPts}
Considering the optimization problem (\ref{eqn:Opt_prob}), the augmented Lagrangian function is given by,
\ifCLASSOPTIONtwocolumn
\begin{equation}\label{eqn:lagr}
\begin{array}{lll}
L(\{\vec{Q}_i\}_{i=1 \text{ to }K}, \{\lambda_i\}_{i=1 \text{ to }K}) &= \sum_{i=1}^K f_i(\{d_i\}|\vec{H}_{i,1},\cdots,\vec{H}_{i,K})\\
& \hspace{8mm} + \sum_{i=1}^K \lambda_i \times (\tau(\vec{Q}_i) - 1)
\end{array},
\end{equation}
\else
\begin{equation}\label{eqn:lagr}
\begin{array}{lll}
L(\{\vec{Q}_i\}_{i=1 \text{ to }K}, \{\lambda_i\}_{i=1 \text{ to }K}) = \sum_{i=1}^K f_i(\{d_i\}|\vec{H}_{i,1},\cdots,\vec{H}_{i,K}) + \sum_{i=1}^K \lambda_i \times (\tau(\vec{Q}_i) - 1)
\end{array},
\end{equation}
\fi
where maximum transmit power at $i^{th}$ transmitter, $P_i$, is assumed to be unity without any loss of generality. There is no assumption on $f_i(\cdot)$, i.e., each Tx-Rx pair can chose any objective function from (\ref{obj:SER}) thro (\ref{obj:MD}) with appropriate sign introduced such that $\sum f_i(\cdot)$ represents minimization problem. 
The gradient of the Lagrangian function (\ref{eqn:lagr}) with respect to the precoder matrix $\vec{Q}_i$ is given by (please refer Appendix~\ref{App:GradComp}),
\ifCLASSOPTIONtwocolumn
\begin{equation}\label{eqn:grad}
\begin{array}{lll}
\triangledown_{Q_i^*}(L) &= \lambda_i \vec{Q}_i -\vec{H}_{i,i}^\text{H} \vec{R}_i^{-1} \vec{H}_{i,i}\vec{Q}_i\vec{E}_i \\
&+ \sum_{l=1,l\neq i}^K \vec{H}_{l,i}^\text{H} \vec{R}_l^{-1} \vec{H}_{l,l} \vec{Q}_l \vec{E}_l\vec{Q}_l^\text{H} \vec{H}_{l,l}^\text{H} \vec{R}_l^{-1} \vec{H}_{l,i} \vec{Q}_i
\end{array},
\end{equation}
\else
\begin{equation}\label{eqn:grad}
\begin{array}{lll}
\triangledown_{Q_i^*}(L) = \lambda_i \vec{Q}_i -\vec{H}_{i,i}^\text{H} \vec{R}_i^{-1} \vec{H}_{i,i}\vec{Q}_i\vec{E}_i + \sum_{l=1,l\neq i}^K \vec{H}_{l,i}^\text{H} \vec{R}_l^{-1} \vec{H}_{l,l} \vec{Q}_l \vec{E}_l\vec{Q}_l^\text{H} \vec{H}_{l,l}^\text{H} \vec{R}_l^{-1} \vec{H}_{l,i} \vec{Q}_i
\end{array},
\end{equation}
\fi
where $\vec{E}_i = \sum_{j_i} \sum_{k_i} \alpha_i^{[j_ik_i]}(\vec{x}_i^{[j_i]} - \vec{x}_i{k_i]}) (\vec{x}_i^{[j_i]} - \vec{x}_i{k_i]})^\text{H}$, and the values of $\alpha_i^{[j_ik_i]}$ for different objective functions are given by,
\ifCLASSOPTIONonecolumn
\begin{equation}
\hspace{-2mm}
\begin{array}{llll}
(\alpha_i^{[jk]})_{MI} = \frac{exp(-d_i^{[jk]})}{\sum_{l} exp(-d_i^{[jl]})}, &
(\alpha_i^{[jk]})_{SER} = exp(-d_i^{[jk]})\\
(\alpha_i^{[jk]})_{BER} = \beta_i^{[jk]} exp(-d_i^{[jk]}), &
(\alpha_i^{[jk]})_{MD} = -r \times (d_i^{[jk]})^{(-r-1)}\\
\end{array}\hspace{-2mm}.
\end{equation}
\else
\begin{gather}
\hspace{-1mm}
\begin{array}{llll}
(\alpha_i^{[jk]})_{MI} = \frac{exp(-d_i^{[jk]})}{\sum_{l} exp(-d_i^{[jl]})}, &\hspace{-2mm}
(\alpha_i^{[jk]})_{SER} = exp(-d_i^{[jk]})\\
(\alpha_i^{[jk]})_{BER} = \beta_i^{[jk]} exp(-d_i^{[jk]}), &\hspace{-2mm}
(\alpha_i^{[jk]})_{MD} = -r \times (d_i^{[jk]})^{(-r-1)}\\
\end{array}\raisetag{2.1\baselineskip}
\end{gather}
\fi

\noindent The choice of the objective function for each Tx-Rx pair is of no consequence while computing the locally optimal point, because the structure of the gradient remains the same irrespective of the objective function (the definition of $\vec{E}_i$ varies for different objective functions, still it represents the covariance matrix of the error vector $(\vec{x}_i^{[j]}-\vec{x}_i^{[k]})$, and hence positive semi-definite). 
The solution for $\{\vec{Q}_i\}_{i=1,\cdots,K}$ (please refer Appendix \ref{App:Prec_struct}) which equates gradient, (\ref{eqn:grad}), to zero is,
\begin{equation}\label{eqn:precStr}
\begin{array}{lll}
\vec{Q}_i = \vec{U}_{H_i} \vec{\varLambda}_{Q_i} \vec{U}_{E_i}^H
\end{array},
\end{equation}

\noindent where $\vec{H}_{i,i}^\text{H} \vec{R}_i^{-1} \vec{H}_{i,i} = \vec{U}_{H_i} \vec{\varLambda}_{H_i} \vec{U}_{H_i}^H$,  $\vec{E}_i = \vec{U}_{E_i} \vec{\varLambda}_{E_i} \vec{U}_{E_i}^H$, and $\vec{U}_{H_i}$ represents the eigen vector matrix, and $\vec{\varLambda}_{H_i}$ represents the diagonal eigen value matrix of $\vec{H}_{i,i}^\text{H} \vec{R}_i^{-1} \vec{H}_{i,i}$, and $\vec{U}_{E_i}$ represents the eigen vector matrix, and $\vec{\varLambda}_{E_i}$ represents the diagonal eigen value matrix of the error covariance matrix $\vec{E}_i$.
The structure of a locally optimal precoder set is the one in which the left singular vector diagonalizes the effective channel matrix, $\vec{H}_{i,i}^\text{H} \vec{R}_i^{-1} \vec{H}_{i,i}$, and the right singular vector diagonalizes the error covariance matrix. The structure given in (\ref{eqn:precStr}) holds regardless of the number of Tx-Rx pairs present in the IC model, and is also valid for the single user MIMO scenario as given in \cite{Xiao2011}. 
Using the structure for the precoder set, it will be shown that IA transmission scheme is a locally optimal point for~(\ref{eqn:Opt_prob}).


\subsection{Interference Alignment} \label{sec:IA_LO}
Interference Alignment scheme for a $K-$user IC was introduced in \cite{Cadambe2008}. The DoF was determined by the dimension of the sub-space occupied by the interfering signal at each receiver, i.e., DoF = min($|\mathcal{S}|$,$M-|\mathcal{I}|$), where $\mathcal{S}$ and $\mathcal{I}$ represent the desired signal and interfering signal subspace at the receiver (assuming $|\mathcal{S}\cup\mathcal{I}|=M$). In a $K-$user IC, if there is no co-operation between the transmitters, $|\mathcal{I}|$ increases with $K$, and the DoF per user decreases with the increase in the number of users. When linear receivers are employed, it was shown in \cite{Cadambe2008} that with precoders designed using alignment of interfering signals, $|\mathcal{I}|$ is independent of $K$, and it takes the minimum value of $\sfrac{M}{2}$. Hence, DoF of $\sfrac{M}{2}$ is achievable. Using Theorem \ref{thm:IAO}, we will show that even for the FA signal set, IA is one of the locally optimal points for the optimization problem (\ref{eqn:Opt_prob}), as SNR$\rightarrow \infty$.
{\theorem For $K-$user MIMO IC, the Interference Alignment transmission scheme in the reciprocal channel is one of the local optima of the optimization problem, when the objective is a function of the distance measure at the receiver.\label{thm:IAO}}

\textit{Proof:} Please refer Appendix \ref{App:thmIAO_pf} and Appendix \ref{App:sec:Pf_OptimalPt_all}.

{\remark The interference in both the forward and the reciprocal channel gets aligned by a precoder and a LMMSE combiner, respectively, only when the noise variance becomes zero. \label{lem:IA}}

From Theorem \ref{thm:IAO},
\ifCLASSOPTIONtwocolumn
\begin{equation}\label{eqn:rank_B_i_rep}
\begin{array}{lll}
\text{rank}([\vec{H}_{1,i}^\text{H} \vec{G_1} \hspace{-1mm}\cdots \vec{H}_{i-1,i}^\text{H} \vec{G_{i-1}}\\
\hspace{15mm}\vec{H}_{i+1,i}^\text{H} \vec{G_{i+1}} \cdots \vec{H}_{K,i}^\text{H} \vec{G_K}])
\end{array}=\frac{M}{2},
\end{equation}
\else
\begin{equation}\label{eqn:rank_B_i_rep}
\begin{array}{lll}
\text{rank}([\vec{H}_{1,i}^\text{H} \vec{G_1} \hspace{-1mm}\cdots \vec{H}_{i-1,i}^\text{H} \vec{G_{i-1}}\quad\vec{H}_{i+1,i}^\text{H}\vec{G_{i+1}} \cdots \vec{H}_{K,i}^\text{H} \vec{G_K}])=\text{rank}(\vec{B}_i)=\frac{M}{2}
\end{array},
\end{equation}
\fi
where $\vec{G}_i = \vec{R}_i^{-1} \vec{H}_{i,i}\vec{Q}_i$ which is the Linear Minimum Mean Square Error (LMMSE) combiner. Theorem \ref{thm:IAO} only talks about the optimality of IA in the reciprocal channel, and does not give any insight into obtaining the optimal precoder. However, when SNR$\rightarrow\infty$, the LMMSE combiner reduces to the Zero Forcing (ZF) combiner. Following \cite{Gomadam2008, Peters2009}, it can be easily argued that aligning the interference in the forward channel is an optimal point as SNR$\rightarrow\infty$.

\subsection{Iterative Algorithm: Sub-Optimal Precoder design}\label{subsec:IAAlgo}
In the previous sub-section (\ref{sec:IA_LO}), it was shown that the conventional IA  scheme is a locally optimal point as SNR$\rightarrow\infty$. Since it was not shown that IA is a globally optimal point, in this sub-section we use an iterative algorithm to investigate the global optimality of the IA precoder design for the EIA scheme.

The objective functions for each Tx-Rx pair can be any one of the utility functions in (\ref{obj:SER}) thro (\ref{obj:MD}), i.e., all the Tx-Rx pair need not consider the same objective function. 
However, for simplicity in the simulations, we use the same objective function at all the Tx-Rx pairs.
A simple CGD algorithm is used to obtain sub-optimal precoders. The algorithm used is the same as Algorithm~1 provided by us in \cite{HariRam2013}, except that the gradient expressions are modified appropriately \footnote{Since the algorithm is similar to \cite{HariRam2013}, for brevity we do not wish to provide the algorithmic steps in this work.}. The detector used at the $i^{th}$ receiver is the MD detector \cite{Kuchi2011}, which is given by,
\begin{equation}\label{eqn:MD_detect}
\begin{array}{llll}
\hat{\vec{x}}_i = \underset{\vec{x}_i\in \mathcal{X}_i}{\operatorname{argmin}} ||\vec{y}_i-\vec{H}_{i,i}\vec{Q}_i \vec{x}_i||_{\vec{R}_i^{-1}}^2
\end{array},
\end{equation}
where $||a||_{\vec{B}}^2$ represents $a^H \vec{B} a$. The MD detector in (\ref{eqn:MD_detect}) decodes the desired signal jointly, but none of the interfering signals are decoded.

\ifCLASSOPTIONtwocolumn
\begin{figure*}
\centering
\begin{minipage}{0.5\textwidth}
\centering
\includegraphics[scale = 0.45]{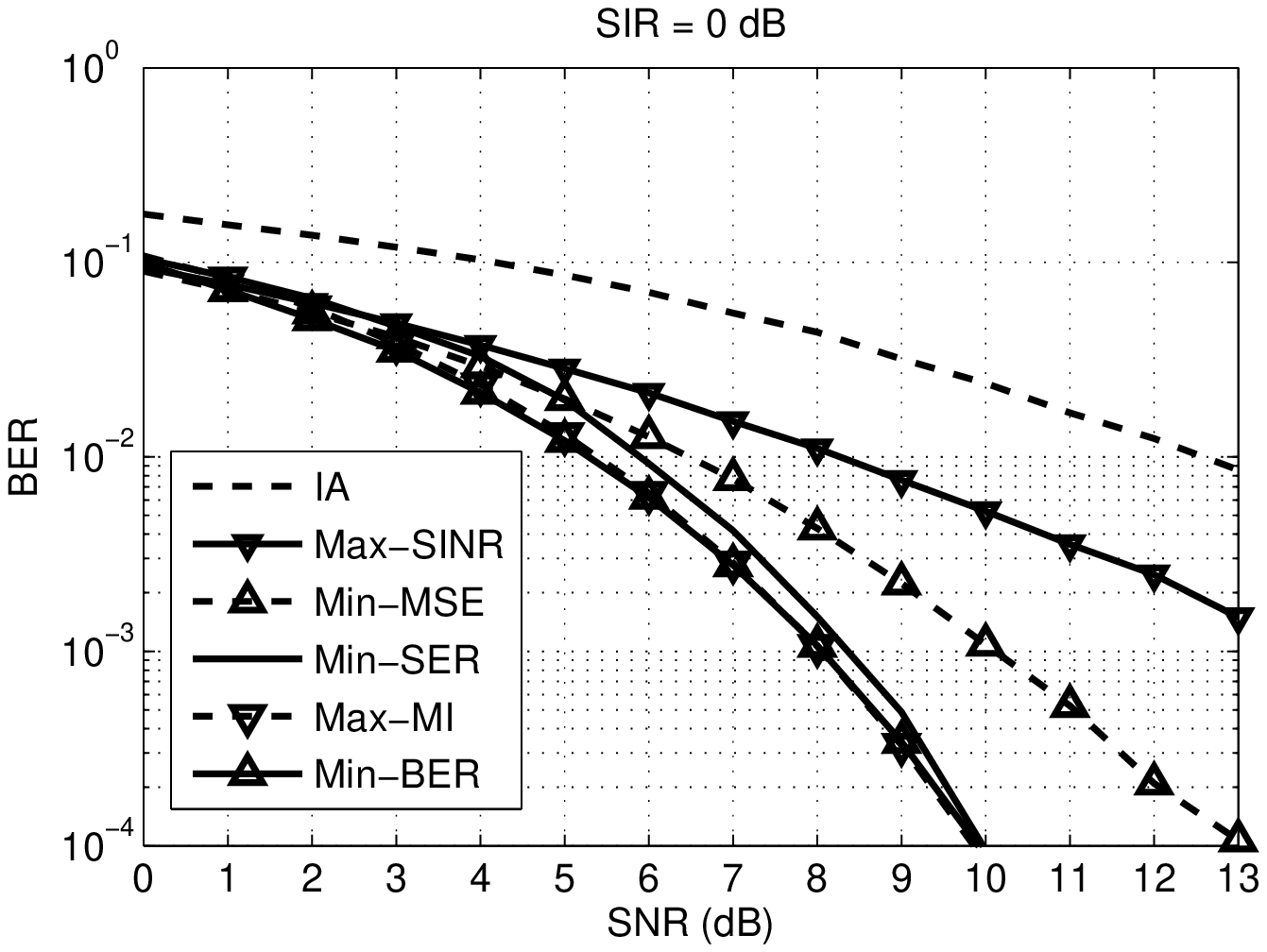}
\caption{Uncoded BER with MD receiver}
\label{fig:Int_Pow_1_unc_ML}
\end{minipage}%
\begin{minipage}{0.5\textwidth}
\centering
\includegraphics[scale = 0.445]{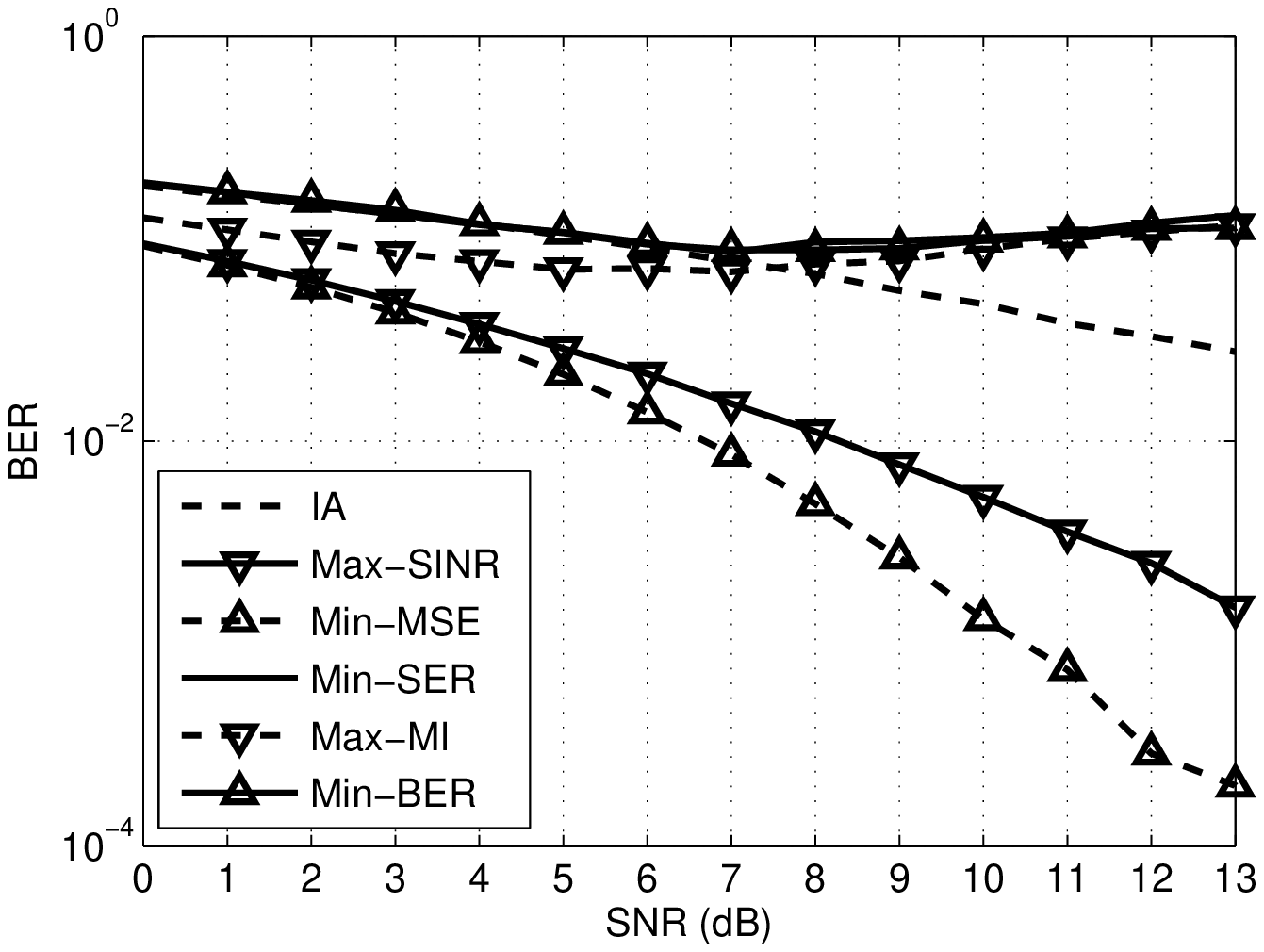}
\caption{Uncoded BER with LMMSE receiver}
\label{fig:Int_Pow_1_unc_LMMSE}
\end{minipage}

\begin{minipage}{0.5\textwidth}
\centering
\includegraphics[scale = 0.45]{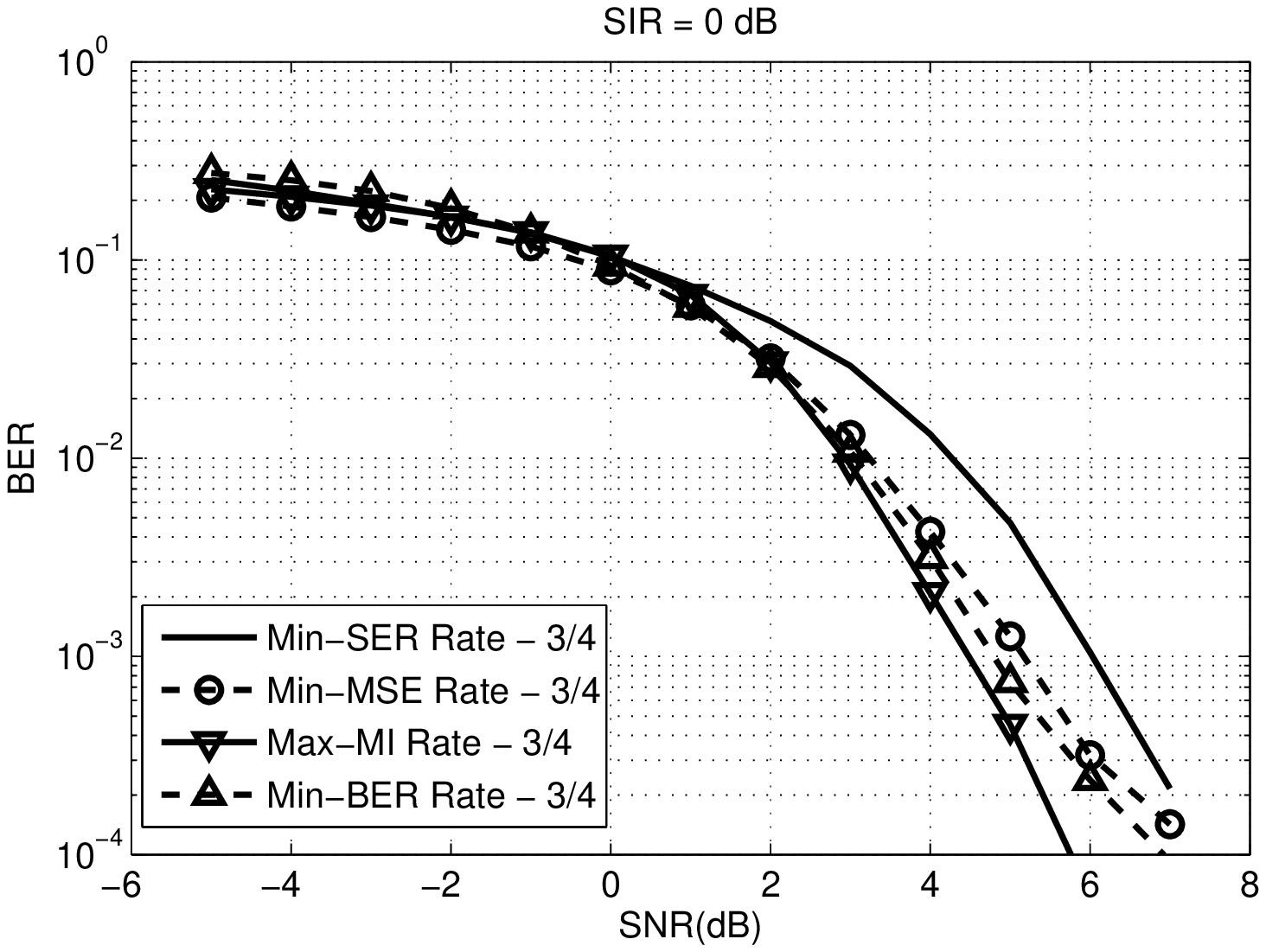}
\caption{Coded BER with code rate = 3/4,\\ \hspace{-38mm} for MD receiver}
\label{fig:Int_Pow_1_c_R34}
\end{minipage}%
\begin{minipage}{0.5\textwidth}
\centering
\includegraphics[scale = 0.45]{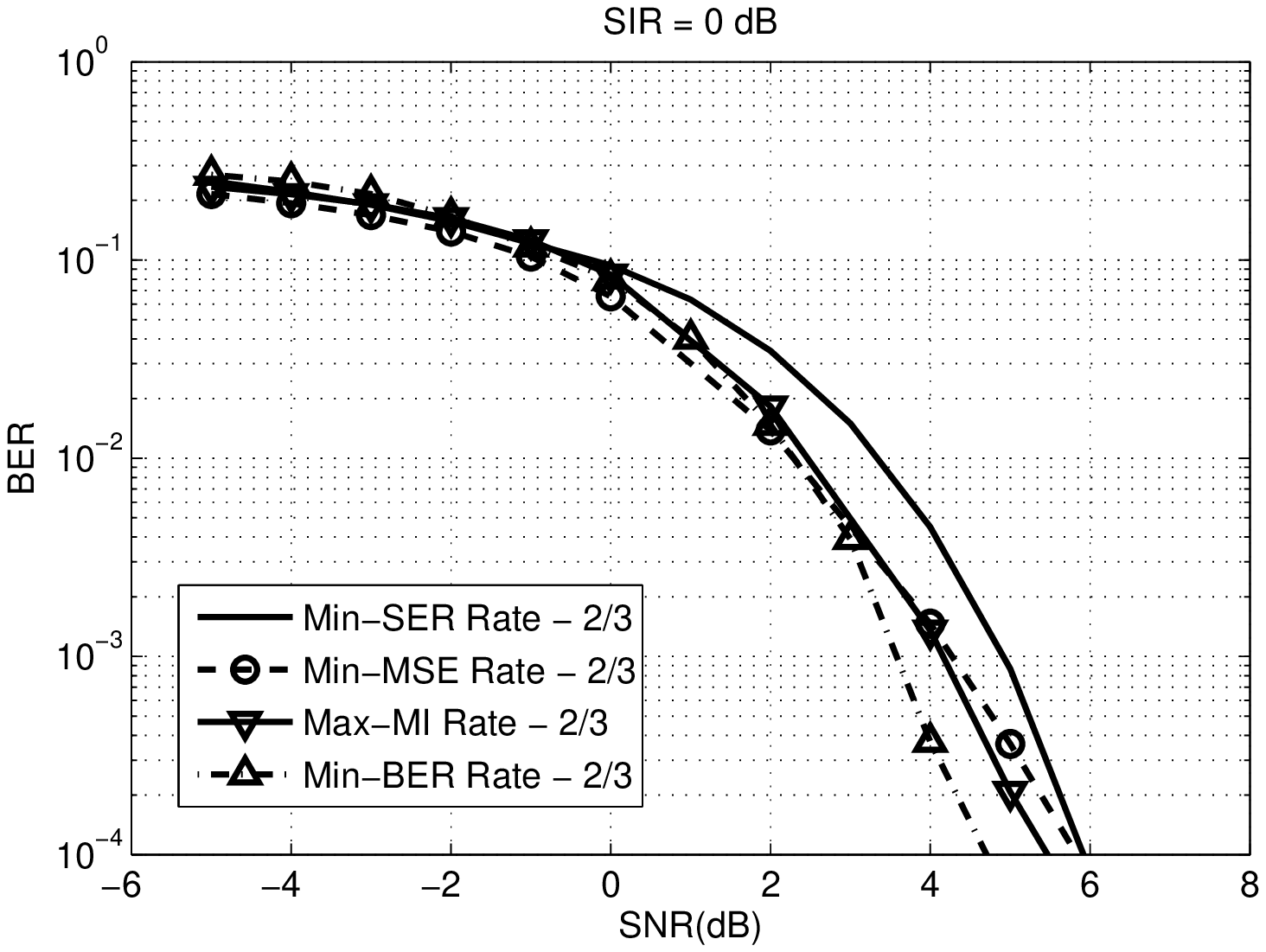}
\caption{Coded BER with code rate = 2/3,\\ \hspace{-38mm} for MD receiver}
\label{fig:Int_Pow_1_c_R23}
\end{minipage}
\end{figure*}
\fi

The considered scenario is a $3-$user MIMO IC. Each transmitter and receiver are equipped with four antennas each, and the FA signals used for transmission are from the QPSK constellation. Two QPSK symbols are transmitted per channel use per transmitter. The channel matrices ($\vec{H}_{i,j}$) are zero mean with covariance given by $E[\vec{H}_{i,j}\vec{H}_{i,j}^\text{H}] = I_M$ and the SIR is unity. 
The SNR is defined as $\sfrac{1}{\sigma^2}$, where $\sigma^2$ represent the noise variance and the maximum transmit power for each transmitter is unity.
The EIA precoder designs are compared with the existing precoder designs provided in \cite{Cadambe2008},\cite{Gomadam2008}, and \cite{Shen2010}.

\ifCLASSOPTIONonecolumn
\begin{figure*}
\centering
\begin{minipage}{0.5\textwidth}
\centering
\includegraphics[scale = 0.5]{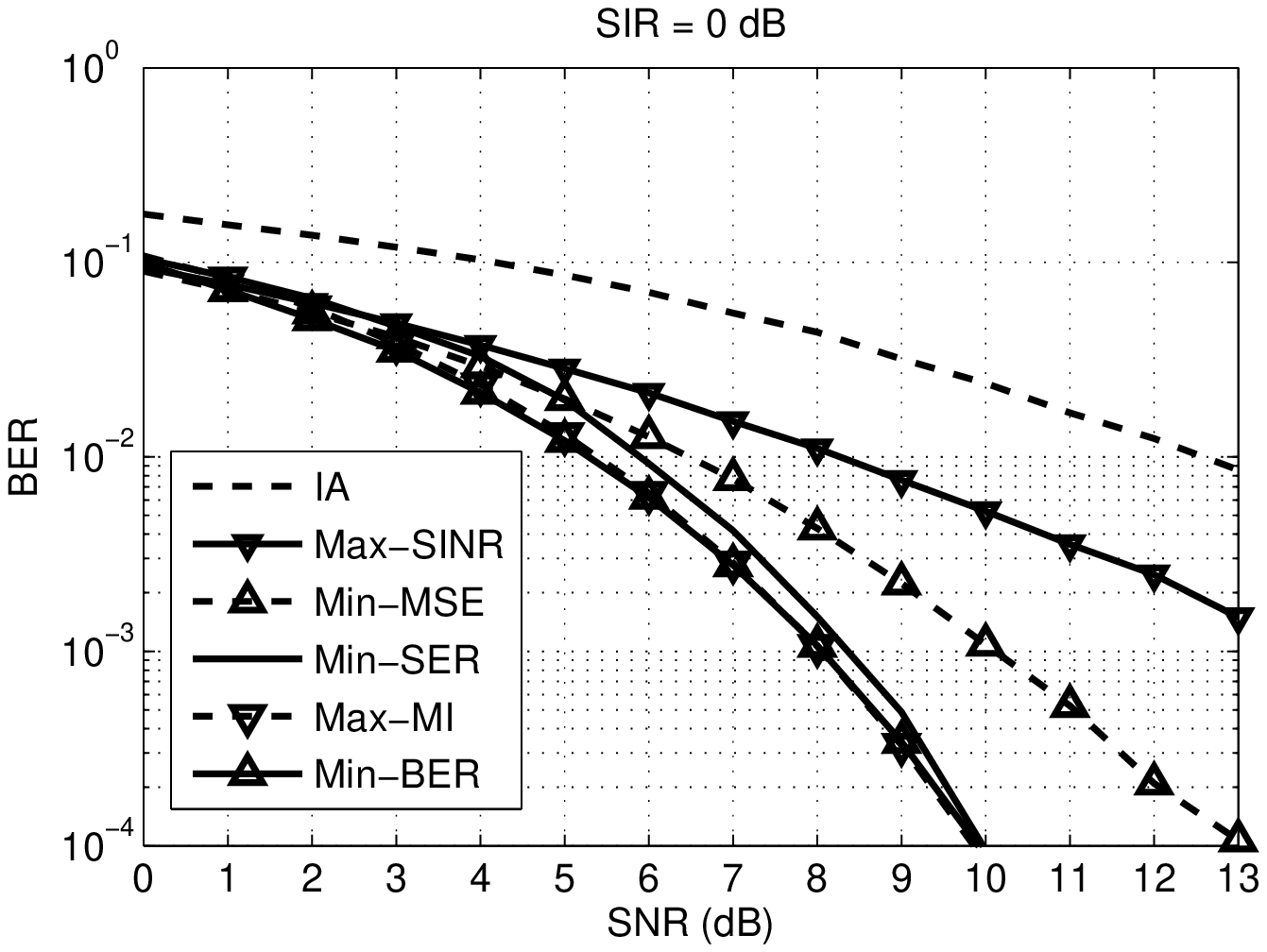}
\caption{Uncoded BER with MD receiver}
\label{fig:Int_Pow_1_unc_ML}
\end{minipage}%
\begin{minipage}{0.5\textwidth}
\centering
\includegraphics[scale = 0.495]{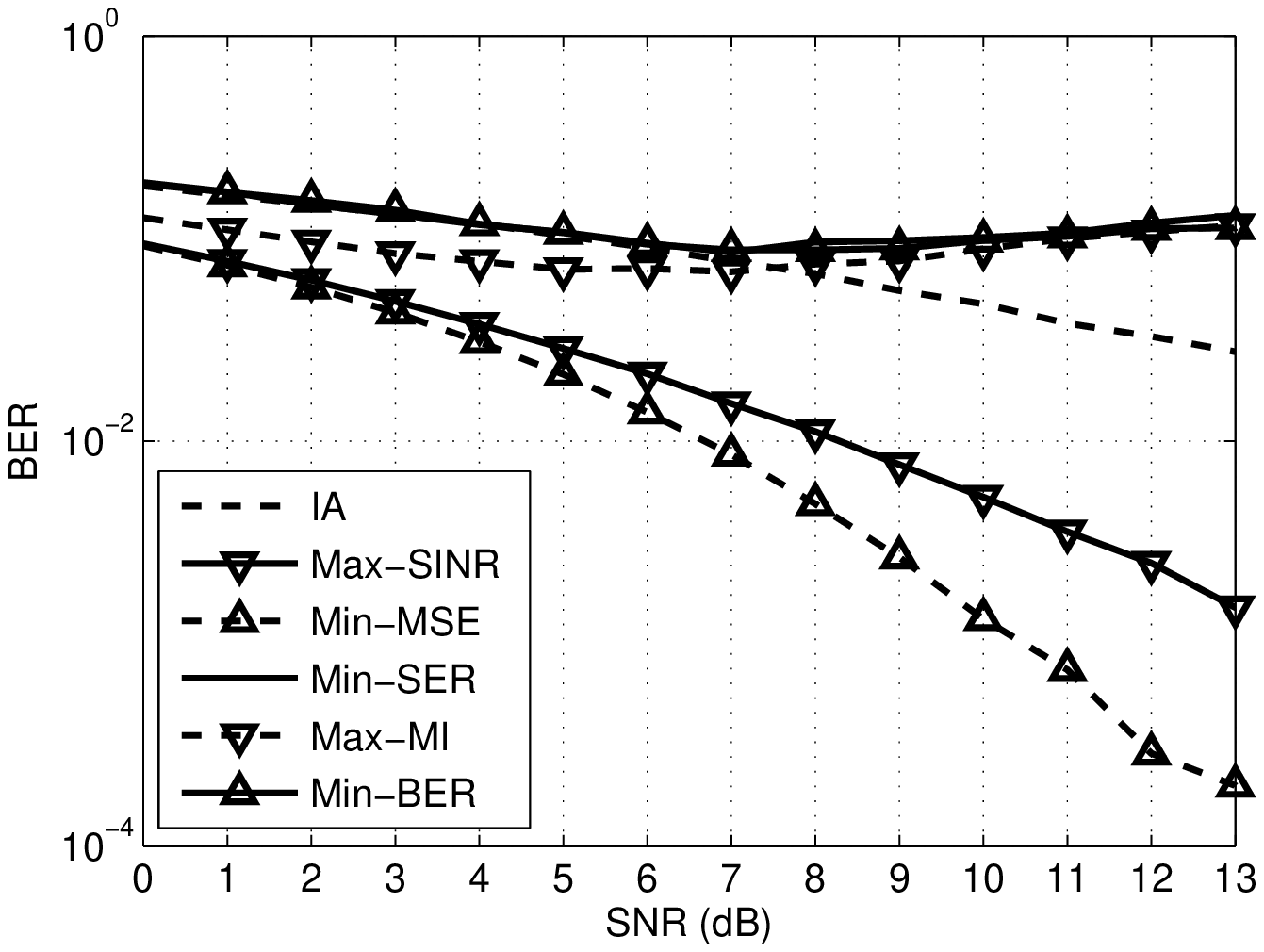}
\caption{Uncoded BER with LMMSE receiver}
\label{fig:Int_Pow_1_unc_LMMSE}
\end{minipage}

\begin{minipage}{0.5\textwidth}
\centering
\includegraphics[scale = 0.45]{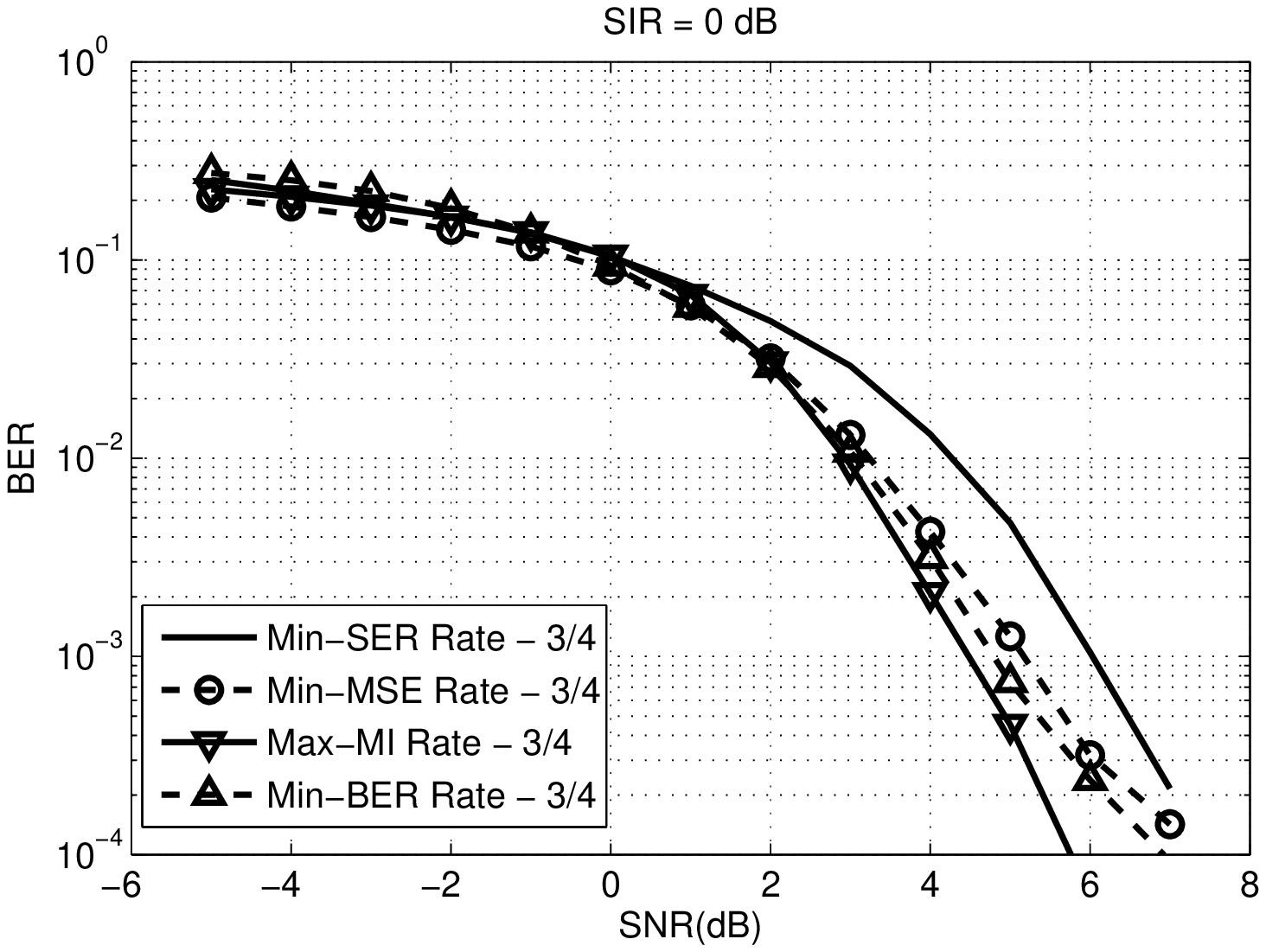}
\caption{Coded BER with code rate = 3/4\\ \vspace{-3mm} \hspace{-45mm} for MD receiver}
\label{fig:Int_Pow_1_c_R34}
\end{minipage}%
\begin{minipage}{0.5\textwidth}
\centering
\includegraphics[scale = 0.45]{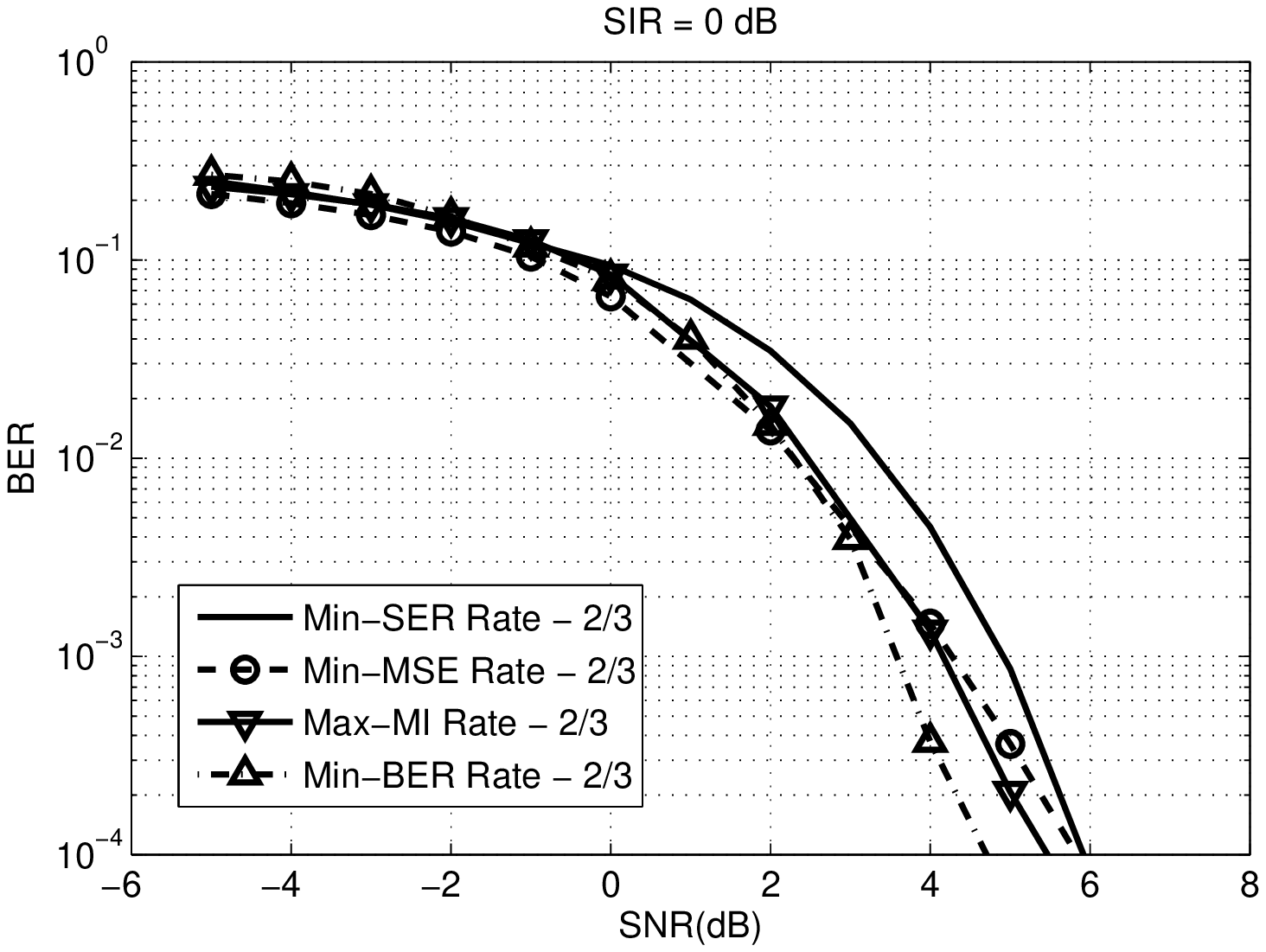}
\caption{Coded BER with code rate = 2/3\\ \vspace{-3mm} \hspace{-45mm} for MD receiver}
\label{fig:Int_Pow_1_c_R23}
\end{minipage}
\end{figure*}
\fi

\noindent The uncoded BER performance is shown in Figs.~ \ref{fig:Int_Pow_1_unc_ML} and~\ref{fig:Int_Pow_1_unc_LMMSE}. Fig. \nolinebreak \ref{fig:Int_Pow_1_unc_ML} shows the BER performance with MD detector for all the algorithms,
where Max-MI represents the optimization problem when $f_i(.)=f_MI(.),\; \forall \;i$, and its similar for other optimization problems.
The gain in SNR to achieve a BER of $10^{-3}$ compared to the Min-MSE is around 2.1 dB for both Max-MI as well as Min-BER schemes. 
When a linear receiver is employed, the BER performance of the optimization problem (\ref{eqn:Opt_prob}) saturates as seen from Fig. \ref{fig:Int_Pow_1_unc_LMMSE} (all the algorithms used LMMSE receiver). This is because of the fact that the interference from the multiple transmitters are not aligned within 2 dimensions ($=\sfrac{M}{2}$). 
Now, the question to be answered is, `\textit{Why is there no error floor in the BER performance when MD detector is used?}'. 
This can be answered using the following Lemma.

\begin{lemma}\label{lemm:MD_const}
In a $M\times M$ MIMO IC, if the interfering signal occupies a $N_i$ dimensional subspace, then the desired signal occupying a subspace of dimension upto $M$ can be decoded with BER$\rightarrow 0$ as SNR$\rightarrow\infty$, if $N_i<M$.
\end{lemma}
\textit{Proof:} Please refer Appendix~\ref{App:sec:lemm_MD_const_proof}.

Since the MD based receiver is able to detect the desired signal even though the interfering signal occupies more than $\sfrac{M}{2}$ dimensions, two questions that remain to be answered are:\\
\textit{
(i) Why does SpAC have to be limited to $\sfrac{1}{2}$? (or) Is it possible to transmit more than $\sfrac{1}{2}$ number of symbols per transmit antenna per channel use, without any error floor happening in the BER performance?\\ (ii) Is it possible to do so without decoding the interferer's data symbols?
}\\
The answers to both these questions is ``\textit{yes}". Using the FIA scheme which is introduced in the next section, it can be shown that any value of SpAC in the range $[0,1]$ can be achieved without incurring any error floor. Also, these values of SpAC can be achieved without decoding any of the interfering signals.

For the sake of completeness, in this work, coded BER performance is also compared for the optimization problem in (\ref{eqn:Opt_prob}) with the Min-MSE algorithm, as in Fig. \ref{fig:Int_Pow_1_c_R34} and Fig. \ref{fig:Int_Pow_1_c_R23}. Since uncoded BER is a function of a distance measure, these objective functions are directly related to uncoded BER. Even then, the coded BER achieves a gain in SNR of about $0.8$~dB at BER=$10^{-3}$ when compared to Min-MSE algorithm. 
The block length of the turbo code is $2048$ for both code rates, the BER curves are averaged over $500$ channel realizatio and a single channel realization is kept constant across $100$ code block. 

\section{Fractional Interference Alignment (FIA)} \label{sec:FIA}
Fractional Interference Alignment transmission scheme is designed under the constraint that all transmitters use finite alphabet signals for transmission. Since FA signal is used, the alignment constraint for the decoding of all the message signals becomes (Lemma~\ref{lemm:MD_const}),
\begin{equation}\label{eqn:FIACons}
\begin{array}{lll}
\mathcal{S}_i^{Rx} \nsubseteq \mathcal{I}_i^{Rx}
\end{array},
\end{equation}
where $\mathcal{S}_i^{Rx}$, and $\mathcal{I}_i^{Rx}$ represents the subspace occupied by the desired signal and interfering signal, respectively, at the $i^{th}$ receiver. 
These subspaces for the IC model (\ref{eqn:sysMod}) are given by,
\ifCLASSOPTIONtwocolumn
\begin{equation}\label{eqn:SandIsubSpa}
\begin{array}{lll}
\mathcal{S}_i^{Rx} &=& \text{span}(\vec{H}_{i,i}\vec{Q}_i)\\
\mathcal{I}_i^{Rx} &=& \operatorname*{\cup}\limits_{j=1, j\neq i}^K \text{ } \text{span}(\vec{H}_{i,j}\vec{Q}_j)
\end{array}.
\end{equation}
\else
\begin{equation}\label{eqn:SandIsubSpa}
\begin{array}{lll}
\mathcal{S}_i^{Rx} = \text{span}(\vec{H}_{i,i}\vec{Q}_i), \quad \text{and} \quad
\mathcal{I}_i^{Rx} = \operatorname*{\cup}\limits_{j=1, j\neq i}^K \text{ } \text{span}(\vec{H}_{i,j}\vec{Q}_j)
\end{array}.
\end{equation}
\fi
The constraint given in (\ref{eqn:FIACons}) can be rewritten as,
\begin{equation}\label{eqn:FIAcons_Equiv}
\begin{array}{lll}
|\mathcal{S}_i^{Rx} \cup \mathcal{I}_i^{Rx}| > |\mathcal{I}_i^{Rx}|, \quad \forall i=1 \text{ to } K
\end{array}.
\end{equation}
Using (\ref{eqn:SandIsubSpa}), and to achieve maximum SpAC, the constraint (\ref{eqn:FIAcons_Equiv}) for the FIA scheme can be rewritten as,
\begin{subequations}\label{eqn:FIAcons_H}
\begin{align}
&|\mathcal{I}_i^{Rx}| < M \label{eqn:FIAcons_H_Inter}\\
\text{and, }& |\mathcal{S}_i^{Rx} \cup \mathcal{I}_i^{Rx}| = M \label{eqn:FIAcons_H_Desir}
\end{align}
\end{subequations}
where $M$ represents the total number of dimension in which signals are received at each receiver. For example in MIMO IC without symbol extension factor, it represents the number of receive antennas. Thus, in the precoder design for FIA scheme, the constraints in (\ref{eqn:FIAcons_H}) is utilized, instead of the linearly independence constraint in \cite{Cadambe2008}. 

\subsection{$3-$user MIMO IC: Without Symbol Extension} \label{subsec:FIA3userMIMO}
The design of FIA precoders is very similar to IA precoder design: (i)~both require global channel knowledge, (ii)~No interfering signals are decoded with perfect alignment of interfering signal within a reduced sub-space, (iii)~No co-operation exists between the receivers while decoding, and finally, (iv)~the designed precoders are linear precoders. Hence, as it can be seen from the proof of Theorem~\ref{thm:FIAMIMO3user}, the design procedure is very similar to \cite{Cadambe2008}, but the constraints are different: (i)~interfering signal are aligned within a sub-space of more than $\sfrac{M}{2}$ dimensions, and (ii)~the desired signal sub-space and interfering signal sub-space are allowed to overlap. In fact, in Theorem \ref{thm:FIAMIMO3user}, we will show that the interfering signals can be confined within a sub-space of dimension $(M-1)$, with SpAC per transmitter of $\sfrac{(M-1)}{M}$.
\ifCLASSOPTIONonecolumn
\vspace{-5mm}
\fi
{\theorem For a 3-user MIMO IC, the maximum achievable SpAC is $\sfrac{(M-1)}{M}$ per transmitter, without an error floor occuring in BER performance as $\sigma^2 \rightarrow 0$, and without any information transfer between the receivers (no co-operation). \label{thm:FIAMIMO3user}}

\textit{Proof:} An example with $M=3$ is provided in Appendix \ref{App:FIA3userMIMO} where precoders are obtained such that (\ref{eqn:FIAcons_H_Inter}) is satisfied. The solution is exactly same as that of \cite{Cadambe2008} for $3-$user MIMO IC alignment expression, except for the precoder dimensions. Hence, the solution for the precoders (with $M>2$) which align the interference from other transmitters is given by,
\ifCLASSOPTIONtwocolumn
\begin{gather}\label{eqn:FIA:IAConst}
\begin{array}{llll}
 \mathbf{Q}_{2} = \mathbf{H}_{1,2}^{-1}\mathbf{H}_{1,3} \mathbf{Q}_{3}; \quad
 \mathbf{Q}_{1} = \mathbf{H}_{2,1}^{-1}\mathbf{H}_{2,3} \mathbf{Q}_{3}; \quad
 \mathbf{Q}_{3} = \Omega \mathbf{T} \mathbf{Q}_{3}\\
  \text{where}, \mathbf{T} = \mathbf{H}_{2,3}^{-1}\mathbf{H}_{2,1} \mathbf{H}_{3,1}^{-1}\mathbf{H}_{3,2} \mathbf{H}_{1,2}^{-1}\mathbf{H}_{1,3}
\end{array}\raisetag{0.8\baselineskip}
\end{gather}
\else
\begin{equation}\label{eqn:FIA:IAConst}
\begin{array}{cccc}
 \mathbf{Q}_{2} = \mathbf{H}_{1,2}^{-1}\mathbf{H}_{1,3} \mathbf{Q}_{3}; \quad
 \mathbf{Q}_{1} = \mathbf{H}_{2,1}^{-1}\mathbf{H}_{2,3} \mathbf{Q}_{3}; \quad
 \mathbf{Q}_{3} = \Omega \mathbf{T} \mathbf{Q}_{3}\\[-0.3cm]
  \text{where}, \mathbf{T} = \mathbf{H}_{2,3}^{-1}\mathbf{H}_{2,1} \mathbf{H}_{3,1}^{-1}\mathbf{H}_{3,2} \mathbf{H}_{1,2}^{-1}\mathbf{H}_{1,3}
\end{array},
\end{equation}
\fi
where $(M-1)$ eigen vectors of the matrix $\vec{T}$ is chosen as the column vectors of the precoder $\vec{Q}_3$. The precoder matrices $\vec{Q}_1$ and $\vec{Q}_2$ can be computed by substituting the value for $\vec{Q}_3$ in \nolinebreak(\ref{eqn:FIA:IAConst}). 
The necessary condition to achieve Zero BER criterion is (\ref{eqn:FIAcons_H_Desir}). Hence, at receiver 1,
\ifCLASSOPTIONtwocolumn
\begin{equation}
\begin{array}{llll}
 \mathcal{S}_i^{Rx} \cup \mathcal{I}_i^{Rx} & \operatorname*{\circeq}& [\vec{H}_{1,1}\vec{Q}_{1} \hspace{0.5cm} \vec{H}_{1,2}\vec{Q}_2 \hspace{0.5cm} \vec{H}_{1,3}\vec{Q}_3]\\[-2mm]
 & \operatorname*{\circeq}\limits_{}^{(a)}& [\vec{H}_{1,1}\vec{Q}_{1}\hspace{0.5cm} \vec{H}_{1,3}\vec{Q}_3]\\[-4mm]
 & \operatorname*{\equiv}\limits_{}^{(b)}& [\vec{W}\vec{Q}_{3}\hspace{0.5cm} \vec{Q}_3]
\end{array},
\end{equation}
\else
\begin{equation}
\begin{array}{llll}
 \mathcal{S}_i^{Rx} \cup \mathcal{I}_i^{Rx} & \operatorname*{\circeq}& [\vec{H}_{1,1}\vec{Q}_{1} \hspace{0.5cm} \vec{H}_{1,2}\vec{Q}_2 \hspace{0.5cm} \vec{H}_{1,3}\vec{Q}_3]\;\;
  \operatorname*{\circeq}\limits_{}^{(a)}& [\vec{H}_{1,1}\vec{Q}_{1}\hspace{0.5cm} \vec{H}_{1,3}\vec{Q}_3]\;\;\operatorname*{\equiv}\limits_{}^{(b)}\;\; [\vec{W}\vec{Q}_{3}\hspace{0.5cm} \vec{Q}_3]
\end{array},
\end{equation}
\fi
where \textit{(a)} is obtained by using the alignment solution (\ref{eqn:FIA:IAConst}) and \textit{(b)} is obtained because the constraint (\ref{eqn:FIAcons_H_Desir}) is on the dimension on the subspace ($\mathcal{S}_i^{Rx} \cup \mathcal{I}_i^{Rx}$), and not on the subspace itself, and, $\vec{W} = \vec{H}_{1,3}^{-1}\vec{H}_{1,1} \vec{H}_{2,1}^{-1} \vec{H}_{2,3}$. The rank of the matrix $[\vec{W}\vec{Q}_{3}\hspace{0.5cm} \vec{Q}_3]$ is $M$ with probability one since all the channel matrices are independent of each other and $\vec{W}$ is a function of $\vec{H}_{1,1}$ while $\vec{Q}_3$ is not a function of $\vec{H}_{1,1}$.

Therefore, the solution given by (\ref{eqn:FIA:IAConst}) satisfies the constraints (\ref{eqn:FIAcons_H}), hence the desired signal can be decoded with the Zero BER criterion. The dimensions of all the precoder matrices are $M \times (M-1)$, and hence the achieved SpAC is $\sfrac{(M-1)}{M}$, which completes the proof for Theorem \ref{thm:FIAMIMO3user}.

\subsection{$3-$user SISO IC: With Symbol Extension} 
\label{subsec:FIA3userSISO}
{\theorem For a $3-$user SISO IC, the maximum achievable SpAC is ($\frac{M-1}{M}$,$\frac{M-1}{M}$,$\frac{M-2}{M}$), without an error floor occuring in BER performance as $\sigma^2 \rightarrow 0$, and without any information transfer between the receivers (no co-operation). \label{thm:FIASISO3user}}

\textit{Proof:} Given a SEF of $M$, the dimension of all the three precoders are fixed as follows: $\vec{Q}_1$ and $\vec{Q}_2$ are $M\times (M-1)$ matrices and $\vec{Q}_3$ is $M \times (M-2)$ matrix, i.e., SpAC for first two transmitters is $\frac{M-1}{M}$ and for the third transmitter SpAC is $\frac{M-2}{M}$. The interfering signal subspace at each receiver is given~by,
\ifCLASSOPTIONtwocolumn
\begin{equation}\label{eqn:IASISO}
\begin{array}{cccc}
\mathcal{I}_1 \circeq [\vec{H}_{1,2}\vec{Q}_{2}\hspace{0.5cm} \vec{H}_{1,3}\vec{Q}_3], \quad \mathcal{I}_2 \circeq [\vec{H}_{2,1}\vec{Q}_{1}\hspace{0.5cm} \vec{H}_{2,3}\vec{Q}_3]\\
\mathcal{I}_3 \circeq [\vec{H}_{3,1}\vec{Q}_{1}\hspace{0.5cm} \vec{H}_{3,2}\vec{Q}_2]
\end{array}.
\end{equation}
\else
\begin{equation}\label{eqn:IASISO}
\begin{array}{cccc}
\mathcal{I}_1 \circeq [\vec{H}_{1,2}\vec{Q}_{2}\hspace{0.5cm} \vec{H}_{1,3}\vec{Q}_3], \quad \mathcal{I}_2 \circeq [\vec{H}_{2,1}\vec{Q}_{1}\hspace{0.5cm} \vec{H}_{2,3}\vec{Q}_3], \quad
\mathcal{I}_3 \circeq [\vec{H}_{3,1}\vec{Q}_{1}\hspace{0.5cm} \vec{H}_{3,2}\vec{Q}_2]
\end{array}.
\end{equation}
\fi
Since the dimension of $\vec{Q}_1$ and $\vec{Q}_2$ is $(M-1)$, the minimum dimension of the interfering signal subspaces ($\mathcal{I}_i$) is ($M-1$). Hence, in order for $|\mathcal{I}_i| = (M-1)$ to happen, the required conditions from (\ref{eqn:IASISO}) are given by,
\begin{equation} \label{eqn:FIASISO_IAsim}
\begin{array}{llll}
\vec{B} \circeq \vec{T} \vec{C}; \text{ }
\vec{A} \subset \vec{B};  \text{ }
\vec{A} \subset \vec{C} \text{ }
\end{array},
\end{equation}
where $\vec{A} = \vec{Q}_3$, $\vec{B} = \vec{H}_{2,3}^{-1} \vec{H}_{2,1} \vec{Q}_1$, $\vec{C} = \vec{H}_{1,3}^{-1} \vec{H}_{1,2} \vec{Q}_2$, and $\vec{T} = \vec{H}_{2,1}^{-1} \vec{H}_{2,3} \vec{H}_{3,1}^{-1} \vec{H}_{3,2} \vec{H}_{1,2}^{-1} \vec{H}_{1,3}$. Note that the alignment in (\ref{eqn:FIASISO_IAsim}), is similar to equations (10)$-$(12) in \cite{Cadambe2008}, except that in (\ref{eqn:FIASISO_IAsim}) the subspace spanned by $\vec{A}$ is smaller than both $\vec{B}$ and $\vec{C}$. Choosing $\vec{A}$, $\vec{B}$, and $\vec{C}$, as,
\ifCLASSOPTIONtwocolumn
\begin{equation}
\begin{array}{llll}
\vec{A} = [\vec{Tw}  &\vec{T}^2 \vec{w} &\cdots &\vec{T}^{M-2}\vec{w}]\\
\vec{B} = [\vec{Tw}  &\vec{T}^2 \vec{w} &\cdots& \vec{T}^{M-1}\vec{w}]\\
\vec{C} = [\vec{w}  &\vec{Tw} &\cdots& \vec{T}^{M-2}\vec{w}]
\end{array},
\end{equation}
\else
\begin{equation}
\begin{array}{llll}
\vec{A} = [\vec{Tw}  \;\;\vec{T}^2 \vec{w} \;\;\cdots \;\;\vec{T}^{M-2}\vec{w}];\;\;\;
\vec{B} = [\vec{Tw}  \;\;\vec{T}^2 \vec{w} \;\;\cdots \;\; \vec{T}^{M-1}\vec{w}];\;\;\;
\vec{C} = [\vec{w}  \;\;\vec{Tw} \;\; \cdots \;\; \vec{T}^{M-2}\vec{w}]
\end{array},
\end{equation}
\fi
we can satisfy the conditions in (\ref{eqn:FIASISO_IAsim}), and the precoders can be obtained from $\vec{A}$, $\vec{B}$, and $\vec{C}$. Here $\vec{w}$ is an arbitrary vector of dimension $M\times 1$. In order for the precoder matrices to satisfy the necessary condition (\ref{eqn:FIAcons_H_Desir}), all the elements of $\vec{w}$ should have non-zero values (using the same procedure as in Appendix \ref{App:FIA3userSISO}). Hence, two transmitters use a SpAC of $\sfrac{(M-1)}{M}$ and $\sfrac{(M-2)}{M}$, which completes the proof for Theorem \ref{thm:FIASISO3user}.

\subsection{$K-$user IC: With Symbol Extension}
\label{subsec:FIAKuserSISO}
\noindent Analytically, finding the precoder for aligning the interference is possible for $3-$user IC. Hence, for $K>3$ asymptotic analysis is performed to obtain the maximum achievable SpAC.

{\theorem The maximum achievable SpAC for a $K-$user SISO IC is {\rm 1} with probability one, and this is achieved as number of dimensions used for transmission tends to infinity.\label{thm:K-useSISO}} 

\textit{Proof:} As considered in \cite{Cadambe2008}, let us assume that all the transmitters use the same signaling subspace $\mathcal{S}$. All the receivers use the same  $\mathcal{S}$ to align the interference. Let the dimension of $\mathcal{S}$ be $M-1$, and let the symbol extension is employed over $M$ symbols. Hence, the dimension of the precoder matrices is $M\times M-1$, and
the alignment expression is given by,
\begin{equation} \label{FIA:Kuser:IA_exp}
\begin{array}{ll}
\vec{H}_{i,j} \vec{Q}_j \circeq \vec{H}_{i,l} \vec{Q}_l
\end{array}, 
\end{equation}
$\forall j,l = 1\cdots K, l\neq j, \forall i = 1 \cdots K$. Let $\vec{Q}_2 = \vec{H}_{3,2}^{-1}\vec{H}_{3,1}\vec{Q}_1$ and $\vec{Q}_i = \vec{H}_{2,i}^{-1}\vec{H}_{2,1}\vec{Q}_1$, for $i=3 \cdots K$. Hence,
\begin{equation} \label{FIA:Kuser:IA_exp_mod}
\begin{array}{ll}
\vec{H}_{i,j} \vec{Q}_j \circeq \vec{T}_{i,j} \vec{Q}_1
\end{array}, 
\end{equation}
where $\vec{T}_{i,1}\rvert_{i = 2 \cdot K} = \vec{H}_{i,1}$,  $\vec{T}_{i,2}\rvert_{i = 1,3 \cdot K} = \vec{H}_{i,2} \vec{H}_{3,2}^{-1}\vec{H}_{3,1}$, and $\vec{T}_{i,j}\rvert_{i = 1 \cdot K, i \neq j} = \vec{H}_{i,j} \vec{H}_{2,i}^{-1}\vec{H}_{2,1}$, $j = 3 \cdots K$. The union of the interference subspace at all the receiver, $\mathcal{I}$, is given by,
\begin{equation}\label{eqn:Kuser_Intsubspace}
\begin{array}{lll}
\mathcal{I} \circeq \vec{Q}_1 \cup \vec{H}_{1,2}\vec{T}_{1,2} \vec{Q}_1 \cdots \cup \vec{T}_{K,K-1} \vec{Q}_1
\end{array},
\end{equation}
which is shown in Fig. \ref{fig:AlignmentEqn}, where $\vec{Q}_1$ is replaced with the subspace $\mathcal{S}$, and is the same alignment expressions as in \cite{Cadambe2008}.

\begin{figure}[h]
\begin{center}
\begin{tikzpicture}[scale=0.9]
    \node (signalSubspace) at (0,1.5) {$\mathcal{S}$};
    \node (intBlock1) at (3,3) [rectangle,draw] {$T_{1,2}$};
    \node (intBlock2) at (3,2) [rectangle,draw] {$T_{1,3}$};
    \node (intBlock3) at (3,0) [rectangle,draw] {$T_{K,K-1}$};
    \node (union) at (5,1.5) [circle,draw] {$\cup$};
    \node (InterferenceSubspace) at (6,1.5) {$\mathcal{I}$};

	\node at (0,1) {Signaling};
	\node at (0,0.6) {Subspace};

	\node at (6,0.9) {Interference};
	\node at (6,0.5) {Subspace};
	
	\draw [->] (signalSubspace) to (1,1.5) --   (2.5,3.5) -- (3.5,3.5) -- (union);
	
	\draw [->] (1,1.5) to (intBlock1.west); 
	\draw [->] (intBlock1.east) to (union);
	
	\draw [->] (1,1.5) to (intBlock2.west); 
	\draw [->] (intBlock2.east) to (union);
	
	\node (vdot) at (3,1) {$\vdots$};
	
	\draw [->] (1,1.5) to (intBlock3.west); 
	\draw [->] (intBlock3.east) to (union);	

    \draw [->] (union) to  (InterferenceSubspace);
\end{tikzpicture}

\end{center}
 \ifCLASSOPTIONonecolumn
\vspace{-10mm}
\fi
\caption{Interference Subspace Construction.}
 \label{fig:AlignmentEqn}
 \ifCLASSOPTIONonecolumn
\vspace{-15mm}
\fi
\end{figure}
In \cite{Cadambe2008}, a construction procedure was given so that asymptotically, as $M\rightarrow \infty$,
\begin{equation}\label{eqn:KuserIC_SandI}
\begin{array}{lll}
\lvert \mathcal{S} \rvert = \lvert \mathcal{I} \rvert
\end{array},
\end{equation}
and from (\ref{eqn:Kuser_Intsubspace}), the subspace occupied by the interferers at all the receivers became equal to the subspace occupied by the signal at each transmitter. This subspace has dimension $M-1$.

At each receiver,
\begin{equation}\label{eqn:FIA_Kuser_SISOneccdn}
\begin{array}{ll}
 [\mathbf{H}_{i,i}\mathcal{S}\hspace{0.5cm} \mathcal{S}]
\end{array},
\end{equation}
and as in section \ref{subsec:FIA3userSISO} the above appended matrix has full rank, namely $M$. Hence, (\ref{eqn:KuserIC_SandI}) and (\ref{eqn:FIA_Kuser_SISOneccdn}) satisfy the constraints in (\ref{eqn:FIAcons_H}), which completes the proof for Theorem \ref{thm:K-useSISO}.

{\corollary The maximum achievable SpAC for a $K-$user MIMO IC is asymptotically {\rm 1} with probability one.}

\textit{Proof:} Treating each transmitter antenna as an independent transmitter, and each receive antenna as an independent receiver, along with the result in Theorem \ref{thm:K-useSISO}, completes this proof.

\section{Discussion}
\subsection{How to achieve different values of SpAC?}
From sections \ref{subsec:FIA3userMIMO} thro \ref{subsec:FIAKuserSISO}, the FIA scheme is used to obtain the maximum SpAC achievable in the IC. 
In order to obtain a value of SpAC which is anywhere between 0 and 1, a subset of the columns of precoder matrices should be chosen such that constraint (\ref{eqn:FIAcons_H}) is satisfied. Since it was shown that the current design satisfies (\ref{eqn:FIAcons_H}), removing some columns of precoder matrix $\vec{Q}_1$ without altering the other precoder matrices will also satisfy (\ref{eqn:FIAcons_H}). Using the same procedure for the other precoder matrices, it can be shown that there exists a set of precoder matrices which satisfies (\ref{eqn:FIAcons_H}), having a value of SpAC between 0 and 1. The constraint used to obtain the IA solution will also satisfy the constraint (\ref{eqn:FIAcons_H}). Hence, the conventional IA solution is one possible way to design the FIA precoders for obtaining $\sfrac{1}{2}$ SpAC.

\subsubsection*{Example}
Consider $M=4$ in a $3-$user IC. Let the eigen vector of $\vec{T}$ (from \ref{eqn:FIA:IAConst}) be given as $\vec{t}_l,\;l=1,\cdots, 4$. To achieve the maximum SpAC of $\sfrac{3}{4}$, chose any three of $\vec{t}_l$ as column vectors of $\vec{Q}_3$ (let $\vec{q}_{3l} = \vec{t}_l,\;l=1,2,3$), and, $\vec{q}_{1l} = \vec{H}_{2,1}^{-1}\vec{H}_{2,3} \vec{q}_{3l}$ and $\vec{q}_{2l} = \vec{H}_{1,2}^{-1}\vec{H}_{1,3} \vec{q}_{3l}$. This precoder set will satisfy the constraint (\ref{eqn:FIACons}), and hence no BER floor will occur.
Having obtained the precoders for $\sfrac{3}{4}$ SpAC, the precoders for $\sfrac{2}{4}$ SpAC can be obtained as follows: $\vec{Q}_i = [\vec{q}_{il}\;\;\vec{q}_{ik}], l,k \in {1,2,3},l\neq k$, since $\vec{t}_4$ is not considered in $\vec{Q}_3$ (and neither are its corresponding vectors  $\vec{H}_{2,1}^{-1}\vec{H}_{2,3} \vec{t}_4$, and $\vec{H}_{1,2}^{-1}\vec{H}_{1,3} \vec{t}_4$ in $\vec{Q}_1$ and $\vec{Q}_2$), the dimension of $\mathcal{I}_i$ can never be $4$. The randomness in $\vec{H}_{i,i}$ ensures that constraint $\mathcal{S}_i \nsubseteq \mathcal{I}_i$ is satisfied for all $i$.  A special case is to select the first and second column vectors of $\vec{Q}_i(\;\forall i)$ from $\sfrac{3}{4}$ SpAC as the column vectors of $\vec{Q}_i$ for $\sfrac{1}{2}$ SpAC, which is equivalent to the design: $\vec{q}_{3l} = \vec{t}_l,\;l=1,2$, and, $\vec{q}_{1l} = \vec{H}_{2,1}^{-1}\vec{H}_{2,3} \vec{q}_{3l}$ and $\vec{q}_{2l} = \vec{H}_{1,2}^{-1}\vec{H}_{1,3} \vec{q}_{3l}$. This is the IA precoder design since $|\mathcal{I}|=2$ and $|\mathcal{S}|+|\mathcal{I}|=4$ are the constraints for IA. 
Hence, for a general SpAC value, there are many ways of designing the precoder from $\vec{Q}_i$'s of $\sfrac{(M-1)}{M}$ SpAC, and IA scheme is one such design with $\sfrac{1}{2}$ SpAC and with no overlap of the desired and interfering signal subspaces.

Also, the FIA solutions obtained are not unique. 
The required constraint for obtaining the precoders is to align the sub-space of interfering signals from multiple transmitters. The non-uniqueness of the FIA solution stems from the fact that post multiplying a matrix by a full rank matrix will not alter the column space \cite{Golub1996}, i.e, span($\vec{AB}$) $\subseteq$ span($\vec{A}$), with $\vec{B}$ being a square matrix and equality if $\vec{B}$ is full rank.

\subsection{What is the expected optimum value of SpAC?} 
Although the solution obtained using the FIA scheme, and the solution provided in \cite{Cadambe2008} are quite similar, there are notable differences between the two schemes as given in Table~\ref{tab:IAvsFIAscheme}.
\begin{table*}
\centering
\ifCLASSOPTIONtwocolumn
\begin{tabular}{|l|c|c|c|}
\else
\scalebox{0.85}{
\begin{tabular}{|l|c|c|c|}
\fi
\hline
&IA scheme & EIA scheme & FIA scheme\\
\hline
\hline
Global Channel knowledge & Required & Required & Required\\[0.1 cm]
\hline
Designed for & Gaussian signals & FA signals & FA signals \\[0.1 cm]
\hline
Receivers used & Linear Receiver & MD detector & MD detector\\[0.1 cm]
\hline
Decoding of Interfering signal & Not required & Not required & Not required\\[0.1 cm]
\hline
Design constraint & $|\mathcal{S} \cup \mathcal{I}| = |\mathcal{S}| + |\mathcal{I}|$ & $-^*$  & $\mathcal{S} \nsubseteq \mathcal{I} $ \\[0.1 cm]
\hline
Overlapping of $\mathcal{S}$ and $\mathcal{I}$ & Not allowed & Allowed & Allowed
\ifCLASSOPTIONtwocolumn
\newline
\fi  (but not necessary)\\[0.1 cm]
\hline
Alignment of Interference & Perfect (within $\sfrac{M}{2}$ dimensions) & $\sfrac{M}{2}\leq |\mathcal{I}|<M$ & Perfect (within
\ifCLASSOPTIONtwocolumn
\newline
\fi $(M-1)$ dimensions)\\[0.1 cm]
\hline
Key parameter & DoF & $-^*$ & SpAC\\[0.1 cm]
\hline
SpAC achieved & $\sfrac{1}{2}$ & $\sfrac{1}{2}$ & $[0,1]$ (Refer Table \ref{tab:OptimumSpAC} 
for optimum value)\\[0.1 cm]
\hline
DoF/dimension &$\sfrac{1}{2}$ & $\sfrac{1}{2}$ &$\sfrac{1}{2}$\\
\hline
\end{tabular}
\ifCLASSOPTIONonecolumn
}
\fi
\ifCLASSOPTIONonecolumn
\caption{Contrast between the FIA and IA schemes. FIA will not provide any benefit over IA\\ \hspace{-10mm} in terms of DoF. For FA signals, the conventional IA is a special case of the FIA design.\\ \hspace{-8mm} Hence, the DoF of FIA is also $\sfrac{1}{2}$. $^*$ The EIA scheme is the conventional IA scheme with\\ \hspace{-20mm} FA signals, and therefore no design parameters are associated with the EIA scheme. } \label{tab:IAvsFIAscheme}
\else
\caption{Contrast between the FIA and IA schemes. FIA will not provide any benefit over IA in terms of DoF. For FA signals, the conventional IA is a special case of the FIA design. Hence, the DoF of FIA is also $\sfrac{1}{2}$, as given. $^*$The EIA scheme is the conventional IA scheme with FA signals, hence we don't associate any design parameter with the EIA scheme. } \label{tab:IAvsFIAscheme}
\fi
\end{table*}

The DoF is of importance only as SNR$\rightarrow\infty$. However, when SNR is finite, it is not desirable to restrict SpAC to $\sfrac{1}{2}$. 
Consider a case when SNR$\rightarrow 0$. In such a scenario, for any value of SIR, it will be desirable to accrue the receive diversity gain\footnote{The intuition is based on the fact that the LMMSE combiner will reduce to the Matched Filter (MF) combiner as SNR$\rightarrow 0$ \cite{Tse2005}, and the MF combiner will try to maximize the receive diversity gain instead of suppressing the interfering signal. The receive diversity gain will be maximum only when receiving a single symbol stream.} present in the channel, instead of trying to attain multiplexing gain. Hence, the expected optimum value of SpAC is $\sfrac{1}{M}$. 
When both SNR and SIR~$\rightarrow\infty$, the $K-$user IC reduces to a single user channel. Hence, the optimum SpAC value tends to $1$. For the zero noise case (SNR$\rightarrow \infty$), SIR is the parameter which limits the performance, and it would be best to avoid the interference, since the proposed FIA scheme does not decode any interfering signals, and treats the interfering signal as colored Gaussian noise. Thus, the expected optimum value of SpAC value will be the same as that of IA scheme ($\sfrac{1}{2}$ SpAC). The FIA scheme results in a broader class of precoder designs which can achieve any value for SpAC from the range $[0,1]$, which covers the full range for SpAC. In general, the optimum value of SpAC ($n^*$, say) depends on both SNR and SIR.
Based on the above arguments we present in the first and third row of table~\ref{tab:OptimumSpAC} the expected values of SpAC of the FIA, as SNR$\rightarrow 0$ and SNR$\rightarrow \infty$, respectively. Further, for finite SNR, the SpAC values of FIA will obviously lie in the same regions defined in the middle row of Table~\ref{tab:OptimumSpAC}, since this is actually the full range of the SpAC values.
\ifCLASSOPTIONonecolumn
\begin{table}[tpb]
\else
\begin{table}[htpb]
\fi
\begin{center}
\ifCLASSOPTIONonecolumn
\scalebox{0.90}{
\fi
\begin{tabular}[b]{|c|c|c|c|}
\hline
  & SIR$\rightarrow 0$ & Finite SIR & SIR$\rightarrow \infty$ \\[0.1 cm]
 \hline
 SNR$\rightarrow 0$ & $n^*\rightarrow \sfrac{1}{M}$ & $n^*\rightarrow \sfrac{1}{M}$ & $n^*\rightarrow \sfrac{1}{M}$ \\[0.1 cm]
\hline
Finite SNR  & $\frac{1}{M} \leq n^* \leq \frac{1}{2}$  & $\frac{1}{M} \leq n^* \leq \frac{(M-1)}{M}$  & $\frac{1}{M} \leq n^* \leq 1$  \\[0.1 cm]
\hline
SNR$\rightarrow \infty$  & $n^* \rightarrow \sfrac{1}{2}$  & $n^*\rightarrow \sfrac{1}{2}$ & $n^*\rightarrow 1$ \\[0.1 cm]
\hline
\end{tabular}
\ifCLASSOPTIONonecolumn
}
\fi
\end{center}
\caption{Expected Optimal number of SpAC per\\ \hspace{-60mm} transmitter}
\label{tab:OptimumSpAC}
\end{table}

\subsection{How to find the optimum value of SpAC?}
In \cite{HariRam2013c}, two optimization problems were used by us to determine the optimal value for SpAC. Since the optimum value of SpAC for minimizing BER will be $\sfrac{1}{M}$ which has the highest diversity gain, while for maximizing MI it will always be $\sfrac{(M-1)}{M}$ which has the highest multiplexing gain, in \cite{HariRam2013c} the following two optimization problems were chosen: (a)~Goodput, and (b)~Coding Gain maximization. Numerical results provided in \cite{HariRam2013c} suggest that for a finite SIR, the value of SpAC increases with SNR, and it saturates at $\sfrac{(M-1)}{M}$. This increase in SpAC with increase in SNR was achieved because the rate of transmission was fixed, and a choice of SpAC $> \sfrac{1}{2}$ was utilized to attain a better BER performance. Thus, from \cite{HariRam2013c}, it can also be argued that the optimum value of SpAC is not only a function of SNR and SIR, but also a function of the objective function used. However, this gain can only be achieved at the expense of increased complexity: (i) in the precoder design, as the design should obtain both, the optimum value of SpAC as well as the precoder sets, and also (ii) in the detector used, as FIA scheme requires a non-linear receiver such as the MD detector. 

\section{Conclusion}
In contrast to the conventional IA scheme which forces the interfering signal to lie within a $\sfrac{M}{2}$ dimensional subspace, we developed a FIA scheme which allows the interference to occupy upto a $(M-1)$ dimensional subspace. This novel precoded transmission scheme was designed to exploit two aspects of the measurement model: (i) the fact that FA signalling schemes are used in practical systems, and (ii) non-linear receiver such as the MD detector is employed to decode the desired signal. 
We have also introduced a ``metric like" quantity called SpAC, representing the number of parallel streams of data signals a node can transmit per channel use. The maximum achievable value for SpAC using the FIA scheme was shown to be unity, and any SpAC value in the full range of $[0,1]$ is achievable by the FIA scheme. Even though FIA will not provide any advantage in terms of DoF, the dimensions of the precoder matrices need not be restricted to $\sfrac{M}{2}$ for any values of SNR when FA signals are used. In other words, unlike the conventional IA for FA signals, or the EIA schemes which restrict the dimension to be half of the number of dimensions available for transmission, FIA allows the precoder dimensions to vary as a function of SNR and SIR. Therefore, FIA provides a better rate or a better BER performance \cite{HariRam2013c}. In this work, we analytically obtained the maximum achievable SpAC value of FIA, which is higher than the SpAC$=\sfrac{1}{2}$ of EIA schemes.

\appendix
\numberwithin{equation}{subsection}
\setcounter{equation}{0}
\subsection{Precoder structure of a locally optimal point for {\rm (\ref{eqn:Opt_prob})}:} \label{App:Prec_struct}
By equating the gradient of (\ref{eqn:Opt_prob}) to zero,
\ifCLASSOPTIONtwocolumn
\begin{equation}\label{eqn:gradZero}
\begin{array}{lll}
\vec{H}_{i,i}^\text{H} \vec{R}_i^{-1} \vec{H}_{i,i}\vec{Q}_i\vec{E}_i = \lambda_i \vec{Q}_i \\
\hspace{15mm}+ \sum_{l=1,l\neq i}^K \vec{H}_{l,i}^\text{H} \vec{R}_l^{-1} \vec{H}_{l,l} \vec{Q}_l \vec{E}_l\vec{Q}_l^\text{H} \vec{H}_{l,l}^\text{H} \vec{R}_l^{-1} \vec{H}_{l,i} \vec{Q}_i
\end{array}.
\end{equation}
\else
\begin{equation}\label{eqn:gradZero}
\begin{array}{lll}
\vec{H}_{i,i}^\text{H} \vec{R}_i^{-1} \vec{H}_{i,i}\vec{Q}_i\vec{E}_i = \lambda_i \vec{Q}_i + \sum_{l=1,l\neq i}^K \vec{H}_{l,i}^\text{H} \vec{R}_l^{-1} \vec{H}_{l,l} \vec{Q}_l \vec{E}_l\vec{Q}_l^\text{H} \vec{H}_{l,l}^\text{H} \vec{R}_l^{-1} \vec{H}_{l,i} \vec{Q}_i
\end{array}.
\end{equation}
\fi
Let $\vec{Q'}_i \vec{\varLambda}_i = \vec{R}_i^{-\frac{1}{2}}\vec{H}_{i,i}\vec{Q}_i\vec{U}_{E_i}$, and $\vec{E}_i = \vec{U}_{E_i} \vec{\varLambda}_{E_i} \vec{U}_{E_i}^H$. Let, 
\begin{equation}\label{eqn:gradVarMani}
\begin{array}{lll}
\vec{A} = \sum_{l=1,l\neq i}^K \vec{H}_{l,i}^\text{H} \vec{R}_l^{-1} \vec{H}_{l,l} \vec{Q}_l \vec{E}_l\vec{Q}_l^\text{H} \vec{H}_{l,l}^\text{H} \vec{R}_l^{-1} \vec{H}_{l,i}
\end{array},
\end{equation}
then (\ref{eqn:gradZero}) can be rewritten as,
\begin{equation} \label{eqn:gradZero_simp}
\begin{array}{ll}
 \vec{Q'}_i \vec{\varLambda}_i \vec{\varLambda}_{E_i} = (\vec{R}_i^{\frac{1}{2}} \vec{H}_{i,i}^{-1\text{H}}(\vec{A}  + \varLambda_i \text{I}_M)\vec{H}_{i,i}^{-1}\vec{R}_i^{\frac{1}{2}})\vec{Q'}_i\vec{\varLambda}_i
\end{array}.
\end{equation}
The above expression is an eigenvector equation, where $\vec{Q'}_i$ represent the eigenvector of the matrix $(\vec{R}_i^{\frac{1}{2}} \vec{H}_{i,i}^{-1\text{H}}(\vec{A}  + \varLambda_i \text{I}_M)\vec{H}_{i,i}^{-1}\vec{R}_i^{\frac{1}{2}})$, which is a Hermitian matrix. Hence, $\vec{Q'}_i$ is a unitary matrix, and from the definition of $\vec{Q'}_i$, the locally optimum solution for (\ref{eqn:Opt_prob}) has the structure given by,
\begin{equation}\label{app:eqn:precStr}
\begin{array}{lll}
\vec{Q}_i = \vec{U}_{H_i} \vec{\varLambda}_{Q_i} \vec{U}_{E_i}^H
\end{array},
\end{equation}
where $\vec{H}_{i,i}^\text{H} \vec{R}_i^{-1} \vec{H}_{i,i} = \vec{U}_{H_i} \vec{\varLambda}_{H_i} \vec{U}_{H_i}^H$. 

\subsection{Proof for Theorem {\rm \ref{thm:IAO}}:}\label{App:thmIAO_pf}
In Appendix~\ref{App:sec:Pf_OptimalPt_all}, it was shown that all extremum points of (\ref{eqn:Opt_prob}) are locally optimum points. Hence, in this section all extremum points are treated as locally optimum points. We will derive a result for $2-$user IC, and extend that result to $K-$user IC with alignment of interfering signals.

For $2-$user IC, substituting the structure of the precoder matrices $\vec{Q}_i$ (\ref{eqn:precStr}), derived in Appendix~\ref{App:Prec_struct}, in 
(\ref{eqn:grad}), and equating it to zero,
we get,
\ifCLASSOPTIONtwocolumn
\begin{equation} \label{opt:2user_Lag_der_rank_0}
\begin{array}{ll}
  \vec{\varLambda}_{Q_1}\vec{\varLambda}_{H_1} \vec{\varLambda}_{E_1} = [ \varLambda_1 \text{I} +
   (\vec{U}_{H_1}^\text{H}\vec{H}_{2,1} \vec{H}_{2,2}^{\text{H}-1}\vec{U}_{H_2})(\vec{\varLambda}_{H_2}\vec{\varLambda}_{Q_2}^2 \vec{\varLambda}_{E_2}\vec{\varLambda}_{H_2})\\
   \hspace{48mm} (\vec{U}_{H_1}^\text{H}\vec{H}_{2,1} \vec{H}_{2,2}^{\text{H}-1}\vec{U}_{H_2})^\text{H}] \vec{\varLambda}_{Q_1}
\end{array}\hspace{-3mm}.
\end{equation}
\else
\begin{equation} \label{opt:2user_Lag_der_rank_0}
\begin{array}{ll}
\hspace{-3mm}  \vec{\varLambda}_{Q_1}\vec{\varLambda}_{H_1} \vec{\varLambda}_{E_1} =& [(\vec{U}_{H_1}^\text{H}\vec{H}_{2,1} \vec{H}_{2,2}^{\text{H}-1}\vec{U}_{H_2})(\vec{\varLambda}_{H_2}\vec{\varLambda}_{Q_2}^2 \vec{\varLambda}_{E_2}\vec{\varLambda}_{H_2}) (\vec{U}_{H_1}^\text{H}\vec{H}_{2,1} \vec{H}_{2,2}^{\text{H}-1}\vec{U}_{H_2})^\text{H} + \varLambda_1 \text{I}] \vec{\varLambda}_{Q_1}
\end{array}.
\end{equation}
\fi

\noindent Let 
$(\vec{U}_{H_1}^\text{H}\vec{H}_{2,1} \vec{H}_{2,2}^{\text{H}-1}\vec{U}_{H_2}) (\vec{\varLambda}_{H_2}^2\vec{\varLambda}_{Q_2}^2 \vec{\varLambda}_{E_2}) (\vec{U}_{H_1}^\text{H}\vec{H}_{2,1} \vec{H}_{2,2}^{\text{H}-1}\vec{U}_{H_2})^\text{H} = \vec{B} = \vec{U}_B\vec{\varLambda}_B \vec{U}_B^\text{H}$, 
where we have used eigen decomposition and $\vec{U}_B$ is a unitary matrix. Here, $\vec{\varLambda}_{H_2}$ and $\vec{\varLambda}_{E_2}$ are full rank i.e., all the diagonal elements are non-zero elements (by definition). Hence, the rank of $\vec{\varLambda}_B$ is purely determined by the rank of the diagonal matrix $\vec{\varLambda}_{Q_2}$ which in turn is the rank of the precoder matrix $\vec{Q}_2$.

Let rank($\vec{\varLambda}_B$) = $1$. This implies that
\begin{equation} \label{opt:2user_Lag_der_rank_1}
\begin{array}{ll}
 \vec{\varLambda}_{Q_1}\vec{\varLambda}_{H_1}\vec{\varLambda}_{E_1} = [\varLambda_B^{[1]} \vec{b}_1\vec{b}_1^\text{H} + \mu_1 \text{I}] \vec{\varLambda}_{Q_1}
\end{array},
\end{equation}
where $\vec{b}_1$ represents the eigen vector corresponding to the non-zero eigen value ($\varLambda_B^{[1]}$). Premultiply (\ref{opt:2user_Lag_der_rank_1}) by $\vec{b}_1^\text{H}$,
\begin{equation} \label{opt:2user_Lag_der_rank_2}
\begin{array}{ll}
  \vec{b}_1^\text{H}[\vec{\varLambda}_{Q_1}\vec{\varLambda}_{H_1}\vec{\varLambda}_{E_1} - (\varLambda_B^{[1]}  + \mu_1 ) \vec{\varLambda}_{Q_1}] = 0
\end{array}.
\end{equation}
For an optimal point to exist, the above equation must be satisfied, i.e., $\vec{b}_1$ must exist. This in turn implies that, rank($\vec{\varLambda}_{Q_1}\vec{\varLambda}_{H_1}\vec{\varLambda}_{E_1} - (\varLambda_B^{[1]} + \mu_1 \text{I}) \vec{\varLambda}_{Q_1}$)$\leq M-1$, and rank($\vec{\varLambda}_{Q_1}$)~=~$M-1$ would satisfy that constraint. Similarly, when rank($\vec{\varLambda}_{Q_2}$)~=~$2$,
\ifCLASSOPTIONtwocolumn
\begin{equation} \label{opt:2user_Lag_der_rank_3}
\begin{array}{ll}
  \vec{b}_1^\text{H}[\vec{\varLambda}_{Q_1}\vec{\varLambda}_{H_1}\vec{\varLambda}_{E_1} - (\varLambda_B^{[1]}  + \mu_1 ) \vec{\varLambda}_{Q_1}] = 0\\
  \vec{b}_2^\text{H}[\vec{\varLambda}_{Q_1}\vec{\varLambda}_{H_1}\vec{\varLambda}_{E_1} - (\varLambda_B^{[2]}  + \mu_1 ) \vec{\varLambda}_{Q_1}] = 0
\end{array},
\end{equation}
\else
\begin{equation} \label{opt:2user_Lag_der_rank_3}
\begin{array}{ll}
  \vec{b}_1^\text{H}[\vec{\varLambda}_{Q_1}\vec{\varLambda}_{H_1}\vec{\varLambda}_{E_1} - (\varLambda_B^{[1]}  + \mu_1 ) \vec{\varLambda}_{Q_1}] = 0,\quad
  \vec{b}_2^\text{H}[\vec{\varLambda}_{Q_1}\vec{\varLambda}_{H_1}\vec{\varLambda}_{E_1} - (\varLambda_B^{[2]}  + \mu_1 ) \vec{\varLambda}_{Q_1}] = 0
\end{array},
\end{equation}
\fi
and rank($\vec{\varLambda}_{Q_1}$) = $M-2$ should satisfy the above equation. And, if the rank of the precoder matrix $\vec{Q}_2$ is $n_2$, then rank($\vec{Q}_1$) = $M-n_2$ is one possible solution. When $n_2=\sfrac{M}{2}$, the resources are equally shared between the two users, and this resource sharing scheme is a locally optimum point. 


For $K-$user IC, let the left singular matrix of the precoder matrix $\vec{Q}_i$ be written as $\vec{U}_{Q_i} = \vec{U}_{Q_i}^{[IA]}\vec{U}_{Q_i}^{[D]}$, where $\vec{U}_{Q_i}^{[IA]}$ is used for interference alignment, and $\vec{U}_{Q_i}^{[D]}$ diagonalizes the effective channel. Now, define $\vec{B}_i$ as follows:
\ifCLASSOPTIONtwocolumn
\begin{equation} \label{opt:3user_B}
\begin{array}{lll}
 \vec{B}_i \hspace{-0mm}&=\hspace{-0mm} \vec{U}_{B_i} \vec{\varLambda}_{B_i} \vec{U}_{B_i}^\text{H}\\
 &=\hspace{-0mm} \sum_{\substack{j=1,\\j\neq i}}^K \vec{H}_{j,i}^\text{H} \vec{R}_j^{-1} \vec{H}_{j,j} \vec{Q}_j \vec{E}_j\vec{Q}_j^\text{H} \vec{H}_{j,j}^\text{H} \vec{R}_j^{-1} \vec{H}_{j,i}\\
 &=\hspace{-0mm} \sum_{\substack{j=1,\\j\neq i}}^K (\vec{H}_{j,i}^\text{H} \vec{R}_j^{-1} \vec{H}_{j,j} \vec{U}_{Q_j}^{[IA]} (\vec{U}_{Q_j}^{[D]} \varLambda_{Q_j} \varLambda_{E_j}\varLambda_{Q_j}^\text{H} \vec{U}_{Q_j}^{[D]\text{H}})\\
& \hspace{45mm}
  \vec{U}_{Q_j}^{[IA]\text{H}} \vec{H}_{j,j}^\text{H} \vec{R}_j^{-1} \vec{H}_{j,i} )
\end{array}.
\end{equation}
\else
\begin{equation} \label{opt:3user_B}
\begin{array}{lll}
 \vec{B}_i \hspace{-0mm}&=&\hspace{-0mm} \vec{U}_{B_i} \vec{\varLambda}_{B_i} \vec{U}_{B_i}^\text{H}
 =\sum_{\substack{j=1,\\j\neq i}}^K \vec{H}_{j,i}^\text{H} \vec{R}_j^{-1} \vec{H}_{j,j} \vec{Q}_j \vec{E}_j\vec{Q}_j^\text{H} \vec{H}_{j,j}^\text{H} \vec{R}_j^{-1} \vec{H}_{j,i}\\
 &=&\hspace{-0mm} \sum_{\substack{j=1,\\j\neq i}}^K \vec{H}_{j,i}^\text{H} \vec{R}_j^{-1} \vec{H}_{j,j} \vec{U}_{Q_j}^{[IA]} (\vec{U}_{Q_j}^{[D]} \varLambda_{Q_j} \varLambda_{E_j}\varLambda_{Q_j}^\text{H} \vec{U}_{Q_j}^{[D]\text{H}})
  \vec{U}_{Q_j}^{[IA]\text{H}} \vec{H}_{j,j}^\text{H} \vec{R}_j^{-1} \vec{H}_{j,i} 
\end{array}.
\end{equation}
\fi
Since $\varLambda_{Q_j}$'s and $\varLambda_{E_j}$'s are full rank, the rank of $\vec{B}_i$ can be given as,
\ifCLASSOPTIONtwocolumn
\begin{equation}\label{eqn:rank_B_i}
\begin{array}{lll}
\hspace{-2mm} \text{rank}(\vec{B}_i)\hspace{-1mm} 
= \hspace{0mm} \text{rank}([\vec{H}_{1,i}^\text{H} \vec{R}_1^{-1} \vec{H}_{1,1} \vec{U}_{Q_1}^{[IA]} \hspace{-1mm}\cdots \vec{H}_{i-1,i}^\text{H} \vec{R}_{i-1}^{-1} \vec{H}_{i-1,i-1} \vec{U}_{Q_{i-1}}^{[IA]}\\
\hspace{15mm}\vec{H}_{i+1,i}^\text{H} \vec{R}_{i+1}^{-1} \vec{H}_{i+1,i+1} \vec{U}_{Q_{i+1}}^{[IA]} \cdots \vec{H}_{K,i}^\text{H} \vec{R}_K^{-1} \vec{H}_{K,K} \vec{U}_{Q_K}^{[IA]}])
\end{array}
\end{equation}
\else
\begin{equation}\label{eqn:rank_B_i}
\begin{array}{lll}
\hspace{-0mm} \text{rank}(\vec{B}_i)\hspace{-0mm} 
= \hspace{0mm} \text{rank}([\vec{H}_{1,i}^\text{H} \vec{R}_1^{-1} \vec{H}_{1,1} \vec{U}_{Q_1}^{[IA]} \hspace{-1mm}\cdots \vec{H}_{i-1,i}^\text{H} \vec{R}_{i-1}^{-1} \vec{H}_{i-1,i-1} \vec{U}_{Q_{i-1}}^{[IA]}\\
\hspace{35mm}\vec{H}_{i+1,i}^\text{H} \vec{R}_{i+1}^{-1} \vec{H}_{i+1,i+1} \vec{U}_{Q_{i+1}}^{[IA]} \cdots \vec{H}_{K,i}^\text{H} \vec{R}_K^{-1} \vec{H}_{K,K} \vec{U}_{Q_K}^{[IA]}])
\end{array}
\end{equation}
\fi
Let $\vec{G}_i$ represent the Linear Minimum Mean Square Error (LMMSE) combiner given by,
\begin{equation} \label{opt:Kuser_IA_LMMSE_receiver}
\begin{array}{lll}
\vec{G}_i &=& \vec{R}_i^{-1} \vec{H}_{i,i} Q_i 
\end{array}.
\end{equation}
Post multiplying a matrix by a full rank matrix will not change the column space \cite{Golub1996}, and using (\ref{opt:Kuser_IA_LMMSE_receiver}), (\ref{eqn:rank_B_i}) can be written~as,
\ifCLASSOPTIONtwocolumn
\begin{gather}\label{eqn:rank_B_i_G_i}
\begin{array}{lll}
\hspace{-2mm} \text{rank}(\vec{B}_i) \\
\hspace{0mm}=  \text{rank}([\vec{H}_{1,i}^\text{H} \vec{G}_1 \cdots \vec{H}_{i-1,i}^\text{H} \vec{G}_{i-1}
\hspace{5mm}\vec{H}_{i+1,i}^\text{H} \vec{G}_{i+1} \cdots \vec{H}_{K,i}^\text{H} \vec{G}_K
])
\end{array}\raisetag{1.5\baselineskip}
\end{gather}
\else
\begin{equation}\label{eqn:rank_B_i_G_i}
\begin{array}{lll}
\hspace{-2mm} \text{rank}(\vec{B}_i) =  \text{rank}([\vec{H}_{1,i}^\text{H} \vec{G}_1 \cdots \vec{H}_{i-1,i}^\text{H} \vec{G}_{i-1}
\hspace{5mm}\vec{H}_{i+1,i}^\text{H} \vec{G}_{i+1} \cdots \vec{H}_{K,i}^\text{H} \vec{G}_K
])
\end{array}.
\end{equation}
\fi
We remark that (\ref{eqn:rank_B_i_G_i}) is the same expression as equation (9) in \cite{Papailiopoulos2012}, when $\vec{G}_j$'s are considered as the precoder matrices and $\vec{H}_{j,i}^\text{H}$'s as the channel matrices. In such a scenario, rank minimization is nothing but the alignment in the reciprocal channel when the LMMSE combiner is used as a precoder. Then, the $K-$user IC reduces to $2-$user IC, with the interfering signal occupying $\sfrac{M}{2}$ dimensional sub-space.
Hence, IA in the reciprocal channel becomes one of the local optimal points, which completes the proof for Theorem \nolinebreak \ref{thm:IAO}.

\subsection{The Extremum point (solution obtained by equating gradient to zero) of {\rm (\ref{eqn:Opt_prob})} is an Optimum point:}\label{App:sec:Pf_OptimalPt_all}
The gradient of the objective function in (\ref{eqn:Opt_prob}) with respect to noise variance ($\sigma^2$) is given by,
\begin{equation} \label{eqn:noiseGrad}
\begin{array}{lll}
 \frac{\partial(C({\vec{Q}_{i}}))}{\partial\sigma^2} = -\sum_{i=1}^{K} Tr(\vec{R}_i^{-1}\vec{H}_{i,i} \vec{Q}_{i} \vec{E}_i\vec{Q}_{i}^\text{H} \vec{H}_{i,i}^\text{H} \vec{R}_i^{-1} )
\end{array},
\end{equation}
where the matrix inside the trace is guaranteed to be positive semi-definite since all the matrices within the trace operation are individually positive definite. Hence, the trace  is a negative value, with the only exception being the case where the precoder is a zero matrix\footnote{When the precoder matrices are zero, MI $=0$, and BER and SER evaluate to one. This setting is the global minima for MI and global maxima for BER and SER, respectively.}. Therefore, $\sfrac{\partial(C({\vec{Q}_{i}}))}{\partial\sigma^2} < 0, \forall \sigma^2$, which implies that 
\ifCLASSOPTIONonecolumn
\begin{equation} \label{eqn:noiseGrad_ineq}
\begin{array}{lll}
 C(\sfrac{\sigma^2}{(1+\epsilon)} | {\vec{Q}_{i}})) > C(\sigma^2 | {\vec{Q}_{i}})) > C(\sfrac{\sigma^2}{(1-\epsilon)} | {\vec{Q}_{i}}))
\end{array},
\end{equation}
\else
\begin{equation} \label{eqn:noiseGrad_ineq}
\begin{array}{lll}
\hspace{-4mm} C(\sfrac{\sigma^2}{(1+\epsilon)} | {\vec{Q}_{i}})) > C(\sigma^2 | {\vec{Q}_{i}})) > C(\sfrac{\sigma^2}{(1-\epsilon)} | {\vec{Q}_{i}}))
\end{array}\hspace{-1mm},
\end{equation}
\fi
for any precoder matrix $\vec{Q}_{i}$.

%

Let $P_i$ represent the power of the precoder, i.e., $\tau(\vec{Q}_{i}) = P_i$. Scaling all the precoders by a factor $\sqrt{\beta}$ can be viewed as transmitting with same precoder but with noise variance being modified to $\sfrac{\sigma^2}{\beta}$. Therefore, from (\ref{eqn:noiseGrad_ineq}), 
\begin{equation} \label{eqn:noiseGrad_pfLocOpt}
\begin{array}{lll}
C(\{\sqrt{(1-\epsilon)}\vec{Q}_i\}) > C(\{\vec{Q}_i\})
\end{array},\text{ for any $\epsilon$ ($>$ 0)}.
\end{equation}
As $\epsilon\rightarrow 0$, $\sqrt{(1-\epsilon)}\vec{Q}_i\rightarrow \vec{Q}_i$, and from the inequality in (\ref{eqn:noiseGrad_pfLocOpt}) the objective function decreases to the extremum points. Hence, all the extremum points of (\ref{eqn:Opt_prob}) are locally optimal points.

\subsection{Proof for Lemma~\ref{lemm:MD_const}:}\label{App:sec:lemm_MD_const_proof}
The proof for Lemma~\ref{lemm:MD_const} proceeds on similar lines to the proof given in \cite{Kuchi2011}, and is given in detail as follows. Consider the system model given by,
\begin{equation}
\begin{array}{lll}
\vec{y} = \vec{H}\vec{Q} \vec{x} + \sum\limits_j \vec{H}_j\vec{Q}_j \vec{x}_j + \vec{n}
\end{array},
\end{equation}
where $\vec{y}$ ($M\times 1$) represents the received signal vector, $\vec{H}$ ($M\times M$) represents the channel between the desired transmitter and receiver, $\vec{Q}$ ($M\times d$) represents the precoder used by the interfering transmitters, $\vec{x}$ ($d\times 1$) represents the symbol vector sent by the desired transmitter. And, $\vec{H}_j$, $\vec{Q}_j$, and $\vec{x}_j$ represent the channel between the $j^{th}$ transmitter and the receiver, precoder matrix used by the $j^{th}$ transmitter, and the symbol vector sent by the $j^{th}$ transmitter, respectively. All the precoder matrices (including $\vec{Q}_j$'s) are assumed to be a function of the channel matrices. 

From \cite{Kuchi2011}, the upper bound on BER for the MD detector is given by,
\begin{equation}
\begin{array}{lll}
P_b \leq \sum\limits_{\vec{x}^{[i]}\in \mathcal{X}} \sum\limits_{\vec{x}^{[j]}\in \mathcal{X}} Q(\sqrt{\vec{e}^{[ij]H} \vec{H}^H \vec{Q}^H \vec{R}^{-1} \vec{H} \vec{Q} \vec{e}^{[ij]}} )
\end{array},
\end{equation}
where $R$ represent the covariance matrix of the interference plus noise, and is given by,
\begin{equation}
\begin{array}{lll}\label{App:eqn:cov_MD_const}
R=\sum_j \vec{H} \vec{Q} (\vec{H} \vec{Q})^H + \sigma^2 \vec{I} = \tilde{\vec{R}}+ \sigma^2 \vec{I} =\vec{U}_{\tilde{R}} (\vec{\varLambda}_{\tilde{R}} + \sigma^2 \vec{I}) \vec{U}_{\tilde{R}}^H
\end{array},
\end{equation}
and, $\vec{e}^{[ij]}=\vec{x}^{[i]}-\vec{x}^{[j]}$.
Using the eigen vector representation of $\vec{R}$ from \eqref{App:eqn:cov_MD_const}, we get
\ifCLASSOPTIONtwocolumn
\begin{equation}
\begin{array}{lll}
d^{[ij]}&=&\vec{e}^{[ij]H} \vec{H}^H \vec{Q}^H \vec{R}^{-1} \vec{H} \vec{Q} \vec{e}^{[ij]}\\
 &=& \vec{e}^{[ij]H} \hat{\vec{H}}^H (\vec{\varLambda}_{\tilde{R}} + \sigma^2\vec{I})^{-1} \hat{\vec{H}} \vec{e}^{[ij]}\\
&=& \sum\limits_m \frac{|\sum_l \hat{h}_{l,m} e_l^{[ij]}|^2}{\lambda_{\tilde{R}}^{[m]} + \sigma^2}
\end{array},
\end{equation}
\else
\begin{equation}
\begin{array}{lll}
d^{[ij]}=\vec{e}^{[ij]H} \vec{H}^H \vec{Q}^H \vec{R}^{-1} \vec{H} \vec{Q} \vec{e}^{[ij]}
 = \vec{e}^{[ij]H} \hat{\vec{H}}^H (\vec{\varLambda}_{\tilde{R}} + \sigma^2\vec{I})^{-1} \hat{\vec{H}} \vec{e}^{[ij]}
= \sum\limits_m \frac{|\sum_l \hat{h}_{l,m} e_l^{[ij]}|^2}{\lambda_{\tilde{R}}^{[m]} + \sigma^2}
\end{array},
\end{equation}
\fi
where $\hat{\vec{H}}= \vec{U}_{\tilde{R}} \vec{H} \vec{Q}$, $e_{l}^{[ij]}$ represent the $l^{th}$ element of $\vec{e}^{[ij]}$, and $\lambda_{\tilde{R}}^{[m]}$ represents the $m^{th}$ diagonal element of $\vec{\varLambda}_{\tilde{R}}$. Now, if $d^{[ij]}\rightarrow~\infty$ then $P_b \rightarrow 0$, and for $d^{[ij]}\rightarrow \infty$ the required conditions are: (a)~SNR($=\sfrac{1}{\sigma^2}$)$\rightarrow \infty$, and (b)~$\lambda_{\tilde{R}}^{[m]}$ must be zero for atleast one value of $m$, i.e., the interference covariance matrix $\tilde{\vec{R}}$ should be rank deficient. Note that there is no constraint on the dimension of the error vector $\vec{e}^{[ij]}$, hence the $\vec{e}^{[ij]}$ or the transmitted symbol vector $\vec{x}$ can be of any dimensions, i.e., even if $d=M$ the receiver can decode the transmitted symbol vector with zero probability of error. It is also apparent that the criteria for $d^{[ij]}\rightarrow \infty$ does not depend on the statistics of the channel, hence the criteria (a) and (b) holds for any channel statistics, and $\mathrm{E}_{H} [P_b] \rightarrow 0$ ($\mathrm{E}_{H}[.]$ represents the expectation over the distribution of the channel matrix $\vec{H}$).

\subsection{Proof for constraint {\rm (\ref{eqn:FIAcons_H_Inter})} in Theorem {\rm \ref{thm:FIAMIMO3user}} for $M=3$:} \label{App:FIA3userMIMO}
Consider a system where all transmitters and receivers are equipped with 3 antennas. It will be shown that each transmitter can transmit 2 symbols each per channel use without causing an error floor at the receiver. The channel matrices ($\mathbf{H}_{i,j}$) are $3\times3$ complex matrices, the precoder matrices ($\mathbf{Q}_i$) are $3\times2$ complex matrices, and the transmitted ($\mathbf{x}_j$) and received ($\mathbf{y}_i$) symbol vectors are $2\times1$ and $3\times1$ complex symbol vectors. 

Let $\mathbf{Q}_i$ = [$\mathbf{q}_{i1}\hspace{0.2cm}\mathbf{q}_{i2}$], where $\mathbf{q}_{ik}$ represents the $k^{th}$ column vector of precoder matrix $\mathbf{Q}_i$. Divide the column vectors into two groups, \{$\mathbf{q}_{11}$,$\mathbf{q}_{21}$,$\mathbf{q}_{31}$\}, and \{$\mathbf{q}_{12}$,$\mathbf{q}_{22}$,$\mathbf{q}_{32}$\}.
For each group, the constraint (\ref{eqn:FIAcons_H_Inter}) is satisfied, if the interference at each receiver is ensured to occupy only one dimension.

The solution for first group of precoder vectors is given as (Appendix IV in \cite{Cadambe2008} for $M=2$),
\ifCLASSOPTIONtwocolumn
\begin{eqnarray}
 \mathbf{q}_{21} = \mathbf{H}_{1,2}^{-1}\mathbf{H}_{1,3} \mathbf{q}_{31} & \quad
 \mathbf{q}_{11} = \mathbf{H}_{2,1}^{-1}\mathbf{H}_{2,3} \mathbf{q}_{31}\label{eqn:rxMLantM12}\\
&\hspace{-35mm} \mathbf{q}_{31} = \alpha \mathbf{T}\mathbf{q}_{31}\label{eqn:rxMLantM13}
\end{eqnarray}
\else
\begin{eqnarray}
 \mathbf{q}_{21} = \mathbf{H}_{1,2}^{-1}\mathbf{H}_{1,3} \mathbf{q}_{31}  \quad
 \mathbf{q}_{11} = \mathbf{H}_{2,1}^{-1}\mathbf{H}_{2,3} \mathbf{q}_{31}\label{eqn:rxMLantM12} \quad
 \mathbf{q}_{31} = \alpha \mathbf{T}\mathbf{q}_{31}
\end{eqnarray}
\fi
where, $\mathbf{T} = \mathbf{H}_{2,3}^{-1}\mathbf{H}_{2,1} \mathbf{H}_{3,1}^{-1}\mathbf{H}_{3,2} \mathbf{H}_{1,2}^{-1}\mathbf{H}_{1,3}$.
Here, $\mathbf{q}_{31}$ is the eigen vector of the matrix $\vec{T}$, and $\mathbf{q}_{11}$ and $\mathbf{q}_{21}$ can be computed from 
(\ref{eqn:rxMLantM12})
. Now, using the solution for the first group of vectors, at each receiver the interference occupies only one dimension from the first group.
Similarly, for the second group of precoders, another eigen vector of $\vec{T}$ is chosen as $\vec{q}_{32}$  ($\vec{q}_{32}\neq \vec{q}_{31}$), and $\mathbf{q}_{12}$, $\mathbf{q}_{22}$ can be computed  from
 (\ref{eqn:rxMLantM12}) using $\vec{q}_{32}$. Thus, at each receiver the interference occupies only one dimension from the second group. Hence, the solution which satisfies (\ref{eqn:FIAcons_H_Inter}) is,
\ifCLASSOPTIONtwocolumn
\begin{gather}\label{eqn:app:IAConst}
\begin{array}{cccc}
 \mathbf{Q}_{2} = \mathbf{H}_{1,2}^{-1}\mathbf{H}_{1,3} \mathbf{Q}_{3}; \quad
 \mathbf{Q}_{1} = \mathbf{H}_{2,1}^{-1}\mathbf{H}_{2,3} \mathbf{Q}_{3}; \quad
 \mathbf{Q}_{3} = \Omega \mathbf{T} \mathbf{Q}_{3}\\
  \text{where}, \mathbf{T} = \mathbf{H}_{2,3}^{-1}\mathbf{H}_{2,1} \mathbf{H}_{3,1}^{-1}\mathbf{H}_{3,2} \mathbf{H}_{1,2}^{-1}\mathbf{H}_{1,3}
\end{array}\raisetag{1\baselineskip}
\end{gather}
\else
\begin{equation}\label{eqn:app:IAConst}
\begin{array}{cccc}
 \mathbf{Q}_{2} = \mathbf{H}_{1,2}^{-1}\mathbf{H}_{1,3} \mathbf{Q}_{3}; \quad
 \mathbf{Q}_{1} = \mathbf{H}_{2,1}^{-1}\mathbf{H}_{2,3} \mathbf{Q}_{3}; \quad
 \mathbf{Q}_{3} = \Omega \mathbf{T} \mathbf{Q}_{3}\\
  \text{where}, \mathbf{T} = \mathbf{H}_{2,3}^{-1}\mathbf{H}_{2,1} \mathbf{H}_{3,1}^{-1}\mathbf{H}_{3,2} \mathbf{H}_{1,2}^{-1}\mathbf{H}_{1,3}
\end{array},
\end{equation}
\fi
where $2$ eigen vectors of the matrix $\vec{T}$ is used as the column vectors for $\vec{Q}_3$, while $\vec{Q}_1$ and $\vec{Q}_2$ are computed from the remaining expressions. Even though, the overall constraint (\ref{eqn:FIACons}) is different from the IA scheme \cite{Cadambe2008}, the alignment constraint (\ref{eqn:FIAcons_H_Inter}) is still the same except for the dimension of the precoder. Hence, the precoders which align the interfering signal within a subspace of dimension strictly less than $M$ is still represented by (\ref{eqn:app:IAConst}).

\subsection{Proof for Theorem {\rm \ref{thm:FIASISO3user}} with $M=3$:} \label{App:FIA3userSISO}
Consider a $3-$user SISO IC where the precoder is applied over 3 symbol duration. It will be shown that when two of the transmitters send 2 symbols while the third transmitter sends only 1 symbol over a 3 symbol duration, the interfering signal at all receiver are confined within a subspace which satisfies (\ref{eqn:FIAcons_H_Inter}). The channel matrices ($\mathbf{H}_{i,j}$) are $3\times3$ complex diagonal matrices, the precoder matrices ($\mathbf{Q}_i$) are $3\times2$ complex matrix for $i=1,2$ and of size $3\times 1$ for $i=3$, and the transmitted ($\mathbf{x}_j$) symbol vector is $2\times 1$ and $1\times 1$ complex symbol vectors for $j=1,2$ and $j=3$ respectively, and the received ($\mathbf{y}_i$) symbol vectors are $3\times1$ complex symbol vectors. 

Let $\mathbf{Q}_i$ = [$\mathbf{q}_{i1}\hspace{0.2cm}\mathbf{q}_{i2}$], for $i=1,2$ and $\vec{Q}_3 = \vec{q}_3$. The interference has to be aligned such that at each receiver interference occupies a subspace of dimension $2(<M)$. The following design procedure satisfies that condition.
At receiver 1,
\begin{equation} \label{FIA:SISO_Rx1}
\begin{array}{lll}
 \vec{H}_{1,2} \vec{q}_{21} = \vec{H}_{1,3} \vec{q}_3
\end{array},
\end{equation}
ensuring the interference occupies only 2 dimension. Similarly at receiver 2 and 3,
\ifCLASSOPTIONtwocolumn
\begin{equation} \label{FIA:SISO_Rx2a3}
\begin{array}{lll}
 \vec{H}_{2,1} \vec{q}_{11} = \vec{H}_{2,3} \vec{q}_3; \quad
 \vec{H}_{3,1} \vec{Q}_1 = \vec{H}_{3,2} \vec{Q}_2
\end{array}.
\end{equation}
\else
\begin{equation} \label{FIA:SISO_Rx2a3}
\begin{array}{lll}
 \vec{H}_{2,1} \vec{q}_{11} = \vec{H}_{2,3} \vec{q}_3; \quad
 \vec{H}_{3,1} \vec{Q}_1 = \vec{H}_{3,2} \vec{Q}_2
\end{array}.
\end{equation}
\fi
Hence, the solution to the above set of equations is,
\ifCLASSOPTIONtwocolumn[-0.3 cm]
\begin{equation} \label{FIA:SISO_Rx2a3_sol}
\begin{array}{lllllll}
\vec{q}_{11} &=& \vec{H}_{2,1}^{-1} \vec{H}_{2,3} \vec{q}_3 & , &
 \vec{q}_{12} &=& \vec{H}_{3,1}^{-1} \vec{H}_{3,2} \vec{q}_{21}\\
 \vec{q}_{21} &=& \vec{H}_{1,2}^{-1} \vec{H}_{1,3} \vec{q}_3 & , &
 \vec{q}_{22} &=& \vec{H}_{3,2}^{-1} \vec{H}_{3,1} \vec{q}_{11}
\end{array}.
\end{equation}
\else
\begin{equation} \label{FIA:SISO_Rx2a3_sol}
\begin{array}{lllllll}
\vec{q}_{11} &=& \vec{H}_{2,1}^{-1} \vec{H}_{2,3} \vec{q}_3 & , &
 \vec{q}_{12} &=& \vec{H}_{3,1}^{-1} \vec{H}_{3,2} \vec{q}_{21}\\
 \vec{q}_{21} &=& \vec{H}_{1,2}^{-1} \vec{H}_{1,3} \vec{q}_3 & , &
 \vec{q}_{22} &=& \vec{H}_{3,2}^{-1} \vec{H}_{3,1} \vec{q}_{11}
\end{array}.
\end{equation}
\fi
The necessary condition to achieve Zero BER criterion is (\ref{eqn:FIAcons_H_Desir}). At receiver 1 and receiver 2, using a similar procedure in section \ref{subsec:FIA3userMIMO} (\ref{eqn:FIAcons_H_Desir}) can be shown trivially, but, at receiver 3,
\ifCLASSOPTIONtwocolumn
\begin{gather}
\begin{array}{llll}
 \mathcal{S}_3^{Rx} \cup \mathcal{I}_3^{Rx} & \operatorname*{\circeq}& [\vec{H}_{3,3}\vec{Q}_{3} \hspace{0.5cm} \vec{H}_{3,1}\vec{Q}_1 \hspace{0.5cm} \vec{H}_{3,2}\vec{Q}_2]\\
 & \operatorname*{\circeq}\limits_{}^{(a)}& [\vec{H}_{3,3}\vec{Q}_{3}\hspace{0.28cm} \vec{W}_{1}\vec{Q}_3 \hspace{0.28cm} \vec{W}_{2}\vec{Q}_3 \hspace{0.28cm} \vec{W}_{3}\vec{Q}_3 \hspace{0.28cm} \vec{W}_{4}\vec{Q}_3]
\end{array}\raisetag{2.1\baselineskip}
\end{gather}
\else
\begin{equation}
\begin{array}{llll}
 \mathcal{S}_3^{Rx} \cup \mathcal{I}_3^{Rx} & \operatorname*{\circeq}& [\vec{H}_{3,3}\vec{Q}_{3} \hspace{0.5cm} \vec{H}_{3,1}\vec{Q}_1 \hspace{0.5cm} \vec{H}_{3,2}\vec{Q}_2]\\
 & \operatorname*{\circeq}\limits_{}^{(a)}& [\vec{H}_{3,3}\vec{Q}_{3}\hspace{0.5cm} \vec{W}_{1}\vec{Q}_3 \hspace{0.5cm} \vec{W}_{2}\vec{Q}_3 \hspace{0.5cm} \vec{W}_{3}\vec{Q}_3 \hspace{0.5cm} \vec{W}_{4}\vec{Q}_3]
\end{array},
\end{equation}
\fi
where, (a) is obtained by using the alignment solution (\ref{FIA:SISO_Rx2a3_sol}). From (\ref{FIA:SISO_Rx2a3_sol}), it should be noted that all the matrices $\vec{W}_i$, for $i=1$ to $4$, is independent of $\vec{H}_{3,3}$. Hence, using the same argument as in $3-$user MIMO IC, the dimension of the subspace $\mathcal{S}_3^{Rx}~\cup~\mathcal{I}_3^{Rx}$ is $M$. However, all the channel matrices are diagonal and hence the the $\vec{W}_i$'s are also diagonal. Hence, in order to satisfy (\ref{eqn:FIAcons_H}) the additional constraint is that none of the elements of $\vec{Q}_3$ ($3\times 1$ vector) can be zero. Thus, in a $3-$user SISO IC, total SpAC = $\sfrac{5}{3}$, as specified by Theorem~\ref{thm:FIASISO3user}, can be achieved provided all the elements of $\vec{Q}_3$ are non-zero, and $\vec{Q}_1$ and $\vec{Q}_2$ are formed using~(\ref{FIA:SISO_Rx2a3_sol}).

\subsection{Gradient Computation:} \label{App:GradComp}
In the computation of gradient the following matrix differentiation identities,\cite{Magnus1988}: 
\ifCLASSOPTIONonecolumn \\ \fi
 $\partial(log(\lvert \vec{X}\rvert)) = Tr(\vec{X}^{-1}\partial(\vec{X}))$, $\partial(\vec{X}^{-1}) = -\vec{X}^{-1}\partial(\vec{X}) \vec{X}^{-1}$, $\partial(Tr(\vec{X})) = Tr(\partial(\vec{X}))$, 
 \ifCLASSOPTIONonecolumn \\ \fi 
 $Tr(\vec{X}^\text{T} \partial(Y^\text{H}))=Tr(\vec{X} \partial(Y^*))$, $Tr(\vec{X}^\text{T} \vec{Y}) = \text{vec}(X)^\text{T} \text{vec}(Y)$ and $\text{vec}(\partial(\vec{X})) = \partial\text{vec}(\vec{X})$, are used. $\partial\text{vec}(\vec{X})$ represents the partial derivative obtained after stacking up all the columns of the matrix into a single column vector.

The derivative of the objective function in (\ref{obj:MI}) will be computed and is generalized for the other objective functions (\ref{obj:SER}) thro (\ref{obj:MD}). Consider the $2-$user IC with objective function given by,

\ifCLASSOPTIONtwocolumn
\begin{gather} \label{obj:MI_2user}
\begin{array}{ll}
C &= -f_{\text{MI}} (\{d_1\}) - f_{\text{MI}} (\{d_2\}) \\
 &= \sum_{j_1} \text{log}_2 \sum_{k_1} exp(-d_1^{[j_1k_1]}) + \sum_{j_2} \text{log}_2 \sum_{k_2} exp(-d_2^{[j_2k_2]}) 
\end{array},\raisetag{2.1\baselineskip} 
\end{gather}
\else
\begin{equation} \label{obj:MI_2user}
\begin{array}{ll}
\hspace{-5mm} C &= -f_{\text{MI}} (\{d_1\}) - f_{\text{MI}} (\{d_2\}) 
 = \sum_{j_1} \text{log}_2 \sum_{k_1} exp(-d_1^{[j_1k_1]}) + \sum_{j_2} \text{log}_2 \sum_{k_2} exp(-d_2^{[j_2k_2]}) 
\end{array},
\end{equation}
\fi
where $d_i^{[jk]}=\vec{e}_i^{[jk]H}\vec{Q}_i^H \vec{H}_{i,i}^H\vec{R}_i^{-1}\vec{H}_{i,i} \vec{Q}_i\vec{e}_i^{[jk]}$, and $\vec{e}_i^{[jk]} = \vec{x}_i^{[j]}-\vec{x}_i^{[k]}$ represents the error vector between $\vec{x}_i^{[j]}$ and $\vec{x}_i^{[k]}$.
The gradient of $C$ with respect to precoder $(\vec{Q}_1^\text{H})$, ($\triangledown_{Q_1^*}(C)$), is computed as follows,
\ifCLASSOPTIONtwocolumn
\begin{equation} \label{obj:MI_der}
\begin{array}{lll}
\partial(C) &=& - \sum_{j_1} \sum_{k_1} \frac{exp(-d_1^{[j_1k_1]})}{\sum_{l_1 exp(-d_1^{[j_1l_1]})}} \partial (d_1^{[j_1k_1]}) \\
&&- \sum_{j_2} \sum_{k_2} \frac{exp(-d_2^{[j_2k_2]})}{\sum_{l_2 exp(-d_2^{[j_2l_2]})}} \partial (d_2^{[j_2k_2]})
\end{array}.
\end{equation}
\else
\begin{equation} \label{obj:MI_der}
\begin{array}{lll}
\partial(C) &=& - \sum_{j_1} \sum_{k_1} \frac{exp(-d_1^{[j_1k_1]})}{\sum_{l_1 exp(-d_1^{[j_1l_1]})}} \partial (d_1^{[j_1k_1]}) - \sum_{j_2} \sum_{k_2} \frac{exp(-d_2^{[j_2k_2]})}{\sum_{l_2 exp(-d_2^{[j_2l_2]})}} \partial (d_2^{[j_2k_2]})
\end{array}.
\end{equation}
\fi

Now, for the first term above,
\ifCLASSOPTIONtwocolumn
\begin{equation} \label{obj:MI_der_des}
\begin{array}{lll}
\partial(d_1^{[j_1k_1]}) &= \partial(Tr(d_1^{[j_1k_1]})) 
= Tr(\partial(d_1^{[j_1k_1]})) \\
&= Tr(\vec{H}_{1,1}^\text{H} \vec{R}_1^{-1} \vec{H}_{1,1}\vec{Q}_1\vec{e}_1^{[j_1k_1]}\vec{e}_1^{[j_1k_1]\text{H}}\hspace{2mm} \partial(\vec{Q}_1^\text{H}))\\
&= \text{vec}\{\vec{H}_{1,1}^\text{H} \vec{R}_1^{-1} \vec{H}_{1,1}\vec{Q}_1 \vec{e}_1^{[j_1k_1]} \vec{e}_1^{[j_1k_1]\text{H}}\}^\text{T} \partial\text{vec}(\vec{Q}_1^*)
\end{array},
\end{equation}
\else
\begin{equation} \label{obj:MI_der_des}
\begin{array}{lll}
\partial(d_1^{[j_1k_1]}) &= \partial(Tr(d_1^{[j_1k_1]})) 
= Tr(\partial(d_1^{[j_1k_1]})) 
= Tr(\vec{H}_{1,1}^\text{H} \vec{R}_1^{-1} \vec{H}_{1,1}\vec{Q}_1\vec{e}_1^{[j_1k_1]}\vec{e}_1^{[j_1k_1]\text{H}}\hspace{2mm} \partial(\vec{Q}_1^\text{H}))\\[-0.3 cm]
&= \text{vec}\{\vec{H}_{1,1}^\text{H} \vec{R}_1^{-1} \vec{H}_{1,1}\vec{Q}_1 \vec{e}_1^{[j_1k_1]} \vec{e}_1^{[j_1k_1]\text{H}}\}^\text{T} \partial\text{vec}(\vec{Q}_1^*)
\end{array},
\end{equation}
\fi
and for the second term in (\ref{obj:MI_der}),
\ifCLASSOPTIONtwocolumn
\begin{equation} \label{obj:MI_der_int}
\begin{array}{lll}
\partial(d_2^{[j_2k_2]})\hspace{-3mm} &= \partial(Tr(d_2^{[j_2k_2]})) \\
&= Tr(\vec{H}_{2,2} \vec{Q}_2 \vec{e}_2^{[j_2k_2]} \vec{e}_2^{[j_2k_2]\text{H}}\vec{Q}_2^\text{H} \vec{H}_{2,2}^\text{H} \hspace{2mm} \partial(\vec{R}_2^{-1}))\\
&= - Tr(\vec{H}_{2,2} \vec{Q}_2 \vec{e}_2^{[j_2k_2]} \vec{e}_2^{[j_2k_2]\text{H}}\vec{Q}_2^\text{H} \vec{H}_{2,2}^\text{H} \vec{R}_2^{-1} \partial(\vec{R}_2) \vec{R}_2^{-1})\\
&= - \text{vec}\{\vec{H}_{2,1}^\text{H} \vec{R}_2^{-1} \vec{H}_{2,2} \vec{Q}_2 \vec{e}_2^{[j_2k_2]} \vec{e}_2^{[j_2k_2]\text{H}}\vec{Q}_2^\text{H} \vec{H}_{2,2}^\text{H}\vec{R}_2^{-1}\\
& \hspace{40mm} \vec{H}_{2,1} \vec{Q}_1\}^\text{T} \hspace{1mm} \partial \text{vec}(\vec{Q}_1^*)
\end{array}\hspace{-2mm}.
\end{equation}
\else
\begin{equation} \label{obj:MI_der_int}
\begin{array}{lll}
\partial(d_2^{[j_2k_2]}) &=& \partial(Tr(d_2^{[j_2k_2]})) 
=
Tr(\vec{H}_{2,2} \vec{Q}_2 \vec{e}_2^{[j_2k_2]} \vec{e}_2^{[j_2k_2]\text{H}}\vec{Q}_2^\text{H} \vec{H}_{2,2}^\text{H} \hspace{2mm} \partial(\vec{R}_2^{-1}))\\
&=& - Tr(\vec{H}_{2,2} \vec{Q}_2 \vec{e}_2^{[j_2k_2]} \vec{e}_2^{[j_2k_2]\text{H}}\vec{Q}_2^\text{H} \vec{H}_{2,2}^\text{H} \vec{R}_2^{-1} \partial(\vec{R}_2) \vec{R}_2^{-1})\\
&=& - \text{vec}\{\vec{H}_{2,1}^\text{H} \vec{R}_2^{-1} \vec{H}_{2,2} \vec{Q}_2 \vec{e}_2^{[j_2k_2]} \vec{e}_2^{[j_2k_2]\text{H}}\vec{Q}_2^\text{H} \vec{H}_{2,2}^\text{H}\vec{R}_2^{-1} \vec{H}_{2,1} \vec{Q}_1\}^\text{T} \hspace{1mm} \partial \text{vec}(\vec{Q}_1^*)
\end{array}\hspace{-2mm}.
\end{equation}
\fi
Using (\ref{obj:MI_der_des}) and (\ref{obj:MI_der_int}) in (\ref{obj:MI_der}), the gradient of $C$ with respect to $\vec{Q}_1$ is given by,
\ifCLASSOPTIONtwocolumn
\begin{equation} \label{obj:MI_der_withSum}
\begin{array}{lll}
\triangledown_{Q_1^*}(C)\hspace{-3mm} &= -\sum_{j_1} \sum_{k_1} \alpha_1^{[j_1k_1]} \vec{H}_{1,1}^\text{H} \vec{R}_1^{-1} \vec{H}_{1,1}\vec{Q}_1\vec{e}_1^{[j_1k_1]} \vec{e}_1^{[j_1k_1]\text{H}}\\
&\hspace{3mm} +\sum_{j_2} \sum_{k_2} (\alpha_2^{[j_2k_2]} \vec{H}_{2,1}^\text{H} \vec{R}_2^{-1} \vec{H}_{2,2} \vec{Q}_2 \vec{e}_2^{[j_2k_2]} \vec{e}_2^{[j_2k_2]\text{H}}\\
&\hspace{40mm}
\vec{Q}_2^\text{H} \vec{H}_{2,2}^\text{H} \vec{R}_2^{-1} \vec{H}_{2,1} \vec{Q}_1)
\end{array},
\end{equation}
\else
\begin{equation} \label{obj:MI_der_withSum}
\begin{array}{lll}
\triangledown_{Q_1^*}(C) &=& -\sum_{j_1} \sum_{k_1} \alpha_1^{[j_1k_1]} \vec{H}_{1,1}^\text{H} \vec{R}_1^{-1} \vec{H}_{1,1}\vec{Q}_1\vec{e}_1^{[j_1k_1]} \vec{e}_1^{[j_1k_1]\text{H}}\\
& &+\sum_{j_2} \sum_{k_2} \alpha_2^{[j_2k_2]} \vec{H}_{2,1}^\text{H} \vec{R}_2^{-1} \vec{H}_{2,2} \vec{Q}_2 \vec{e}_2^{[j_2k_2]} \vec{e}_2^{[j_2k_2]\text{H}}
\vec{Q}_2^\text{H} \vec{H}_{2,2}^\text{H} \vec{R}_2^{-1} \vec{H}_{2,1} \vec{Q}_1
\end{array},
\end{equation}
\fi
where, $\alpha_1^{[j_1k_1]} = \frac{exp(-d_1^{[j_1k_1]})}{\sum_{l_1} exp(-d_1^{[j_1l_1]})}$ and, $\alpha_2^{[j_2k_2]} = \frac{exp(-d_2^{[j_2k_2]})}{\sum_{l_2} exp(-d_2^{[j_2l_2]})}$.
Define $\vec{E}_1$ and $\vec{E}_2$ as the error covariance matrix, $\vec{E}_i = \sum_{j_i} \sum_{k_i} \alpha_i^{[j_ik_i]}\vec{e}_i^{[j_ik_i]} \vec{e}_i^{[j_ik_i]\text{H}}$, $i=1,2$. The gradient can then be expressed as,
\ifCLASSOPTIONtwocolumn
\begin{gather} \label{obj:MI_der_2user}
\begin{array}{lll}
\triangledown_{Q_1^*}(C) =& -\vec{H}_{1,1}^\text{H} \vec{R}_1^{-1} \vec{H}_{1,1}\vec{Q}_1\vec{E}_1 \\
&+ \vec{H}_{2,1}^\text{H} \vec{R}_2^{-1} \vec{H}_{2,2} \vec{Q}_2 \vec{E}_2\vec{Q}_2^\text{H} \vec{H}_{2,2}^\text{H} \vec{R}_2^{-1} \vec{H}_{2,1} \vec{Q}_1
\end{array}\raisetag{2.1\baselineskip}
\end{gather}
\else
\begin{equation} \label{obj:MI_der_2user}
\begin{array}{lll}
\triangledown_{Q_1^*}(C) =& -\vec{H}_{1,1}^\text{H} \vec{R}_1^{-1} \vec{H}_{1,1}\vec{Q}_1\vec{E}_1 + \vec{H}_{2,1}^\text{H} \vec{R}_2^{-1} \vec{H}_{2,2} \vec{Q}_2 \vec{E}_2\vec{Q}_2^\text{H} \vec{H}_{2,2}^\text{H} \vec{R}_2^{-1} \vec{H}_{2,1} \vec{Q}_1
\end{array}.
\end{equation}
\fi
Similarly, the gradient with respect to the precoder $\vec{Q}_i$ for a $K$-user IC is given by,
\ifCLASSOPTIONtwocolumn
\begin{gather} \label{obj:MI_der_kuser}
\begin{array}{lll}
\triangledown_{Q_i^*}(C) =& -\vec{H}_{i,i}^\text{H} \vec{R}_i^{-1} \vec{H}_{i,i}\vec{Q}_i\vec{E}_i \\
&+ \sum_{l=1,l\neq i}^K \vec{H}_{l,i}^\text{H} \vec{R}_l^{-1} \vec{H}_{l,l} \vec{Q}_l \vec{E}_l\vec{Q}_l^\text{H} \vec{H}_{l,l}^\text{H} \vec{R}_l^{-1} \vec{H}_{l,i} \vec{Q}_i
\end{array}\raisetag{2.1\baselineskip}
\end{gather}
\else
\begin{equation} \label{obj:MI_der_kuser}
\begin{array}{lll}
\triangledown_{Q_i^*}(C) = -\vec{H}_{i,i}^\text{H} \vec{R}_i^{-1} \vec{H}_{i,i}\vec{Q}_i\vec{E}_i + \sum_{l=1,l\neq i}^K \vec{H}_{l,i}^\text{H} \vec{R}_l^{-1} \vec{H}_{l,l} \vec{Q}_l \vec{E}_l\vec{Q}_l^\text{H} \vec{H}_{l,l}^\text{H} \vec{R}_l^{-1} \vec{H}_{l,i} \vec{Q}_i
\end{array}.
\end{equation}
\fi
Now considering the objective from (\ref{obj:SER}) thro (\ref{obj:MD}) will result in exactly same expression as in \ref{obj:MI_der_kuser}, except for the change in the variable $\alpha_i^{[jk]}$ in the definition of the error covariance matrix $\vec{E}_i$. The values for $\alpha_i^{[jk]}$ are given by,
\ifCLASSOPTIONonecolumn
\begin{equation}
\hspace{-2mm}
\begin{array}{llll}
(\alpha_i^{[jk]})_{MI} = \frac{exp(-d_i^{[jk]})}{\sum_{l} exp(-d_i^{[jl]})}, &
(\alpha_i^{[jk]})_{SER} = exp(-d_i^{[jk]})\\
(\alpha_i^{[jk]})_{BER} = \beta_i^{[jk]} exp(-d_i^{[jk]}), &
(\alpha_i^{[jk]})_{MD} = -r \times (d_i^{[jk]})^{(-r-1)}\\
\end{array}\hspace{-2mm}.
\end{equation}
\else
\begin{gather}
\hspace{-1mm}
\begin{array}{llll}
(\alpha_i^{[jk]})_{MI} = \frac{exp(-d_i^{[jk]})}{\sum_{l} exp(-d_i^{[jl]})}, &\hspace{-2mm}
(\alpha_i^{[jk]})_{SER} = exp(-d_i^{[jk]})\\
(\alpha_i^{[jk]})_{BER} = \beta_i^{[jk]} exp(-d_i^{[jk]}), &\hspace{-2mm}
(\alpha_i^{[jk]})_{MD} = -r \times (d_i^{[jk]})^{(-r-1)}\\
\end{array}\raisetag{2.1\baselineskip}
\end{gather}
\fi



\begin{thebibliography}{10}
\baselineskip 12pt
\providecommand{\url}[1]{#1}
\csname url@samestyle\endcsname
\providecommand{\newblock}{\relax}
\providecommand{\bibinfo}[2]{#2}
\providecommand{\BIBentrySTDinterwordspacing}{\spaceskip=0pt\relax}
\providecommand{\BIBentryALTinterwordstretchfactor}{4}
\providecommand{\BIBentryALTinterwordspacing}{\spaceskip=\fontdimen2\font plus
\BIBentryALTinterwordstretchfactor\fontdimen3\font minus
  \fontdimen4\font\relax}
\providecommand{\BIBforeignlanguage}[2]{{%
\expandafter\ifx\csname l@#1\endcsname\relax
\typeout{** WARNING: IEEEtran.bst: No hyphenation pattern has been}%
\typeout{** loaded for the language `#1'. Using the pattern for}%
\typeout{** the default language instead.}%
\else
\language=\csname l@#1\endcsname
\fi
#2}}
\providecommand{\BIBdecl}{\relax}
\BIBdecl

\bibitem{Shannon1961}
C.~E. Shannon, ``{Two-way communication channels},'' in \emph{Berkely Symp. on
  Mathematical Statistics and Probability}, vol.~1.\hskip 1em plus 0.5em minus
  0.4em\relax Univ. California Press, 1961, pp. 611--644.

\bibitem{Ahlswede1974}
R.~Ahlswede, ``{The capacity region of a channel with two senders and two
  receivers},'' \emph{The Annals of Probability}, vol.~2, no.~5, pp. 805--814,
  1974.

\bibitem{Carleial1978}
A.~Carleial, ``{Interference channels},'' \emph{{IEEE} Trans. Inform. Theory},
  vol.~24, no.~1, pp. 60--70, Jan 1978.

\bibitem{Sato1981}
H.~Sato, ``{The capacity of the Gaussian interference channel under strong
  interference (Corresp.)},'' \emph{{IEEE} Trans. Inform. Theory}, vol.~27,
  no.~6, pp. 786--788, Nov 1981.

\bibitem{Han1981}
T.~Han and K.~Kobayashi, ``{A new achievable rate region for the interference
  channel},'' \emph{{IEEE} Trans. Inform. Theory}, vol.~27, no.~1, pp. 49--60,
  Jan 1981.

\bibitem{Gamal1982}
A.~Gamal and M.~Costa, ``{The capacity region of a class of deterministic
  interference channels (Corresp.)},'' \emph{{IEEE} Trans. Inform. Theory},
  vol.~28, no.~2, pp. 343--346, Mar 1982.

\bibitem{Costa1985}
M.~H.~M. Costa, ``{On the Gaussian interference channel},'' \emph{{IEEE} Trans.
  Inform. Theory}, vol.~31, no.~5, pp. 607--615, Sep 1985.

\bibitem{Etkin2008}
R.~Etkin, D.~Tse, and H.~Wang, ``{Gaussian Interference Channel Capacity to
  Within One Bit},'' \emph{{IEEE} Trans. Inform. Theory}, vol.~54, no.~12, pp.
  5534--5562, Dec 2008.

\bibitem{Cadambe2008}
V.~Cadambe and S.~Jafar, ``{Interference Alignment and Degrees of Freedom of
  the $K-$User Interference Channel},'' \emph{{IEEE} Trans. Inform. Theory},
  vol.~54, no.~8, pp. 3425--3441, Aug 2008.

\bibitem{Gomadam2008}
K.~Gomadam, V.~Cadambe, and S.~Jafar, ``{Approaching the Capacity of Wireless
  Networks through Distributed Interference Alignment},'' in \emph{IEEE Global
  Telecommunications Conference}, Dec 2008, pp. 1--6.

\bibitem{Peters2009}
S.~Peters and R.~Heath, ``{Interference alignment via alternating
  minimization},'' in \emph{IEEE International Conference on Acoustics, Speech
  and Signal Processing (ICASSP)}, Apr 2009, pp. 2445--2448.

\bibitem{Shen2010}
H.~Shen, B.~Li, M.~Tao, and Y.~Luo, ``{The New Interference Alignment Scheme
  for the MIMO Interference Channel},'' in \emph{IEEE Wireless Communications
  and Networking Conference}, April 2010, pp. 1--6.

\bibitem{Schmidt2009}
D.~Schmidt, C.~Shi, R.~Berry, M.~Honig, and W.~Utschick, ``{Minimum Mean
  Squared Error interference alignment},'' in \emph{Asilomar Conference on
  Signals, Systems and Computers}, Nov 2009, pp. 1106--1110.

\bibitem{Papailiopoulos2012}
D.~Papailiopoulos and A.~Dimakis, ``{Interference Alignment as a Rank
  Constrained Rank Minimization},'' \emph{{IEEE} Trans. Signal Processing},
  vol.~60, no.~8, pp. 4278--4288, Aug 2012.

\bibitem{HariRam2013}
B.~{Hari Ram}, W.~Li, A.~Ayyar, J.~Lilleberg, and K.~Giridhar, ``{Precoder
  Design for $K-$User Interference Channels with Finite Alphabet Signals},''
  \emph{IEEE Communications Letters}, vol.~17, no.~4, pp. 681--684, 2013.

\bibitem{Fadlallah2013}
Y.~Fadlallah, A.~Khandani, K.~Amis, A.~A{\"\i}ssa-El-Bey, and R.~Pyndiah,
  ``{Precoding and Decoding in the MIMO Interference Channel for Discrete
  Constellation},'' in \emph{International Symposium on Personal Indoor and
  Mobile Radio Communications (PIMRC)}, 2013.

\bibitem{Willems2010}
\BIBentryALTinterwordspacing
F.~Willems, ``{Information and Communication Theory: Communication Theory},''
  2010. [Online]. Available:
  \url{http://www.sps.ele.tue.nl/members/F.M.J.Willems/TEACHING_files/5JK00/infcomtheory2010.pdf}
\BIBentrySTDinterwordspacing

\bibitem{Kuchi2011}
K.~Kuchi and A.~Ayyar, ``{Performance Analysis of ML Detection in MIMO Systems
  with Co-Channel Interference},'' \emph{{IEEE} Commun. Lett.}, vol.~15, no.~8,
  pp. 786--788, Aug 2011.

\bibitem{Yetis2009}
C.~Yetis, T.~Gou, S.~Jafar, and A.~Kayran, ``{Feasibility Conditions for
  Interference Alignment},'' in \emph{IEEE Global Telecommunications
  Conference}, Dec 2009, pp. 1--6.

\bibitem{HariRam2013c}
B.~{Hari Ram} and K.~{Giridhar}, ``{Precoder Design for Fractional Interference
  Alignment},'' in \emph{{Forty Seventh Asilomar Conference on Signals, Systems
  and Computers}}, Nov 2013.

\bibitem{Huang2011}
H.~Huang and V.~Lau, ``{Partial Interference Alignment for $K-$User MIMO
  Interference Channels},'' \emph{{IEEE} Trans. Signal Processing}, vol.~59,
  no.~10, pp. 4900--4908, Oct 2011.

\bibitem{Wu2013}
Y.~Wu, C.~Xiao, X.~Gao, J.~Matyjas, and Z.~Ding, ``{Linear Precoder Design for
  MIMO Interference Channels with Finite-Alphabet Signaling},'' \emph{IEEE
  Transactions on Communications}, vol.~61, no.~9, pp. 3766--3780, 2013.

\bibitem{Ganesan2012}
A.~Ganesan and B.~S. Rajan, ``{On Precoding for Constant K-User MIMO Gaussian
  Interference Channel with Finite Constellation Inputs},'' \emph{arXiv
  preprint arXiv:1210.3819}, 2012.

\bibitem{Cioffi2007a}
\BIBentryALTinterwordspacing
J.~Cioffi, ``{Course notes for digital communication: Signal processing},''
  2007. [Online]. Available:
  \url{http://www.stanford.edu/group/cioffi/book/chap1.pdf}
\BIBentrySTDinterwordspacing

\bibitem{Moon2000}
T.~Moon and W.~Stirling, \emph{{Mathematical methods and algorithms for signal
  processing}}.\hskip 1em plus 0.5em minus 0.4em\relax Prentice hall New York,
  2000, vol.~1.

\bibitem{Chiani1997}
M.~Chiani, ``{Analytical distribution of linearly modulated cochannel
  interferers},'' \emph{{IEEE} Trans. Commun.}, vol.~45, no.~1, pp. 73--79, Jan
  1997.

\bibitem{Xiao2011}
C.~Xiao, Y.~Zheng, and Z.~Ding, ``{Globally Optimal Linear Precoders for Finite
  Alphabet Signals Over Complex Vector Gaussian Channels},'' \emph{{IEEE}
  Trans. Signal Processing}, vol.~59, no.~7, pp. 3301--3314, July 2011.

\bibitem{Golub1996}
G.~Golub and C.~{Van Loan}, \emph{{Matrix computations}}.\hskip 1em plus 0.5em
  minus 0.4em\relax Johns Hopkins University Press, 1996.

\bibitem{Tse2005}
D.~Tse and P.~Viswanath, \emph{{Fundamentals of wireless communication}}.\hskip
  1em plus 0.5em minus 0.4em\relax Cambridge Univ Pr, 2005.

\bibitem{Magnus1988}
J.~Magnus and H.~Neudecker, \emph{{Matrix differential calculus with
  applications in statistics and econometrics}}.\hskip 1em plus 0.5em minus
  0.4em\relax John Wiley \& Sons, 1988.

\end{thebibliography}
\end{document}